\documentstyle[pre,aps,multicol,eqsecnum,graphicx,amssymb]{revtex}
\newcommand{\bleq}{\ifpreprintsty \else
\end{multicols}\vspace*{-3.5ex}{\tiny \noindent\begin{tabular}[t]{c|}
\parbox{0.493\hsize}{~} \\ \hline \end{tabular}} \fi}
\newcommand{\eleq}{\ifpreprintsty \else
{\tiny\hspace*{\fill}\begin{tabular}[t]{|c}\hline
\parbox{0.49\hsize}{~} \\
\end{tabular}}\vspace*{-2.5ex}\begin{multicols}{2} \fi}
\newcommand{\bcols}{\ifpreprintsty\else\begin{multicols}{2}\fi}
\newcommand{\ecols}{\ifpreprintsty\else\end{multicols}\fi}
\newcommand{\be}{\begin{eqnarray}} \newcommand{\ee}{\end{eqnarray}}

\begin{document}
\draft

\title{Fluctuating ``Pulled'' Fronts: the Origin and the Effects of a
Finite Particle Cutoff}

\author{Debabrata Panja and Wim van Saarloos}

\maketitle

\begin{center}
{\em Instituut--Lorentz, Universiteit Leiden, Postbus 9506, 2300 RA
Leiden, The Netherlands}
\end{center}

\date{\today}

\begin{abstract} 
Recently it has been shown that when an equation that allows so-called
pulled fronts in the mean-field limit is modelled with a stochastic
model with a finite number $N$ of particles per correlation volume,
the convergence to the speed $v^*$ for $N \to \infty$ is extremely
slow --- going only as $\ln^{-2}N$. Pulled fronts are fronts that
propagate into an unstable state, and the asymptotic front speed is
equal to the linear spreading speed $v^*$ of small linear
perturbations about the unstable state. In this paper, we study the
front propagation in a simple stochastic lattice model. A detailed
analysis of the microscopic picture of the front dynamics shows that
for the description of the far tip of the front, one has to abandon
the idea of a uniformly translating front solution. The lattice and
finite particle effects lead to a ``stop-and-go'' type dynamics at the
far tip of the front, while the average front behind it ``crosses
over'' to a uniformly translating solution. In this formulation, the
effect of stochasticity on the asymptotic front speed is coded in the
probability distribution of the times required for the advancement of
the ``foremost bin''. We derive expressions of these probability
distributions by matching the solution of the far tip with the
uniformly translating solution behind. This matching includes various
correlation effects in a mean-field type approximation. Our results
for the probability distributions compare well to the results of
stochastic numerical simulations. This approach also allows us to deal
with much smaller values of $N$ than it is required to have the
$\ln^{-2}N$ asymptotics to be valid. Furthermore, we show that if one
insists on using a uniformly translating solution for the entire front
ignoring its breakdown at the far tip, then one can obtain a simple
expression for the corrections to the front speed for finite values of
$N$, in which various subdominant contributions have a clear physical
interpretation.

\end{abstract} 


\bcols

\section{Introduction}
\subsection{Fronts and fluctuation effects}

In pattern forming systems quite often situations occur where  patches
of different bulk phases occur which are separated by fronts or
interfaces. In such cases, the relevant dynamics is usually dominated
by the dynamics of these fronts. When the interface separates two
thermodynamically stable phases, as in crystal-melt interfacial growth
problems, the width of the interfacial zone is usually of atomic
dimensions. For such systems, one often has to resort to a moving
boundary description in which the boundary conditions at the interface
are determined phenomenologically or by microscopic considerations. A
question that naturally arises for such interfaces is the influence of
stochastic fluctuations on the motion and scaling properties of such
interfaces.

At the other extreme is a class of fronts that arise in systems that
form patterns, and in which the occurrence of fronts or transition
zones is fundamentally related to their nonequilibrium nature, as they
do not connect two thermodynamic equilibrium phases which are
separated by a first order phase transition.  In such cases --- for
example, chemical fronts \cite{meron}, the temperature and density
transition zones in thermal plumes \cite{tabeling}, the domain walls
separating domains of different orientation in  in rotating
Rayleigh-B\'enard convection \cite{tucross}, or streamer fronts in
discharges \cite{streamers} --- the fronts are relatively wide and
therefore described by the same continuum equations that describe
nonequilibrium bulk patterns. The lore in nonequilibrium pattern
formation is that when the relevant length scales are large, (thermal)
fluctuation effects are relatively small \cite{ch}. For this reason,
the dynamics of many pattern forming systems can be understood in
terms of the deterministic dynamics of the basic patterns and coherent
structures. For fronts,  the first questions to study are  therefore
properties like existence and  speed of propagation of the front
solutions of the deterministic equations, which in most cases are
partial differential equations.  In the last decades, the fundamental
propagation mechanism of such deterministic fronts has become
relatively well understood.

From the above perspective, it is maybe less of a surprise that the
detailed questions concerning the stochastic properties of inherently
nonequilibrium fronts have been addressed, to some extent, only
relatively recently
\cite{breuer,lemar,armeroprl,bd,kns,levine,armero,riordan}, and that
it has taken a while for researchers to become fully aware of the fact
that the so-called pulled fronts \cite{dee,bj,vs2,ebert} which
propagate into an unstable state, do  {\em not} fit into the common
mold: they have anomalous sensitivity to particle effects
\cite{bd,kns,levine} , and have been argued to display uncommon
scaling behavior \cite{riordan,rocco,tripathy1,tripathy2}.

{\em Pulled fronts} are fronts which propagate into an unstable state,
and whose propagation dynamics is essentially that they are being
``pulled along'' by the growth and spreading of the small
perturbations about the unstable state into which the front propagates
--- their asymptotic speed $v_{as}$ is equal to the linear spreading
speed $v^*$ of perturbations about the unstable state: $v_{as}=v^*$
\cite{dee,bj,vs2,ebert}. This contrasts with the {\em pushed fronts},
for  which $v_{as} > v^*$, and whose dynamics is determined by the
nonlinearities in the dynamical equations \cite{bj,vs2,ebert}. The
behavior of pushed fronts is essentially similar to fronts between two
(meta)stable states.

The concept of a pulled front most  naturally fits a formulation of
the dynamical equations in terms of {\em continuum variables}, for by
``small perturbations'' we mean that the deviations of the field
values from the values in the unstable state are small enough that
nonlinear terms in the deviations can be neglected. From various
directions, it has become clear in the last few years that such pulled
fronts do show very unusual behavior and response to perturbations.
First of all, Brunet and Derrida have shown that when the continuum
field equations are used for a finite particle model so as to have a
growth cutoff at the field value $1/N$, where $N$ is the typical
number of particles in the bulk phase behind the front, the deviation
from the continuum value $v^*$ of the front speed is often large, and
it vanishes only as $1/\ln^2N$ (with a known prefactor which they
calculated) \cite{bd}. On the other hand, we recently found that with
an infinitesimal growth cutoff and a similarly infinitesimal growth
{\it enhancement\/} behind it, one can have a much higher front speed
than $v^*$ \cite{PvS}. Furthermore, the scaling properties of pulled
fronts in stochastic field equations with a particular type of
multiplicative noise have been found to be anomalous: in  one
dimension, they  are predicted to exhibit subdiffusive wandering
\cite{rocco}, but in higher dimensions their scaling behavior is given
by the KPZ equation \cite{kpz} in one dimension higher than one would
naively expect \cite{tripathy1,tripathy2} (the question to what extent
these results are applicable to lattice models, where the finite
particle effects always make the fronts weakly pushed, is still a
matter of debate \cite{notepulled,moro}).  Moreover, even without
fluctuations, pulled fronts respond differently to coupling to other
fields, e.g., they never reduce to standard moving boundary problems,
even if they are thin \cite{ebert2}.

All these effects have one origin in common, namely the fact that the
dynamics of pulled fronts, by its very nature, is not determined by
the nonlinear front region itself, but by the region {\em at the
leading edge of the front},  where deviations from the unstable state
are small. To a large  degree, this semi-infinite region alone
determines the  universal relaxation of the  speed of a deterministic
pulled front to its asymptotic value \cite{bd,vs2,ebert}, as well as
the  anomalous scaling behavior of stochastic fronts
\cite{rocco,tripathy1,tripathy2,notepulled,moro} in continuum
equations with multiplicative noise. As realized by Brunet and Derrida
\cite{bd}, the crucial importance of the region, where the deviations
from  the unstable state are small, also implies that if one builds a
lattice model version of a front propagating into an unstable state,
the front speed is surprisingly sensitive to the dynamics of the tip
(the far end) of the front where only one or a few particles per
lattice site are present. It is this effect which is the main subject
of this paper.

\subsection{Open questions \label{IB}}
If we study fronts for a field describing the number density $\phi$ of
particles, and normalize the field in such a way that its average
value  behind the front, where there are $N$ particles per unit of
length, is 1, then at the very far end of the leading edge, where the
discrete particle nature of the actual model becomes most noticeable,
the value of the normalized number density field is of order $1/N$.
Brunet and Derrida \cite{bd} therefore modeled the effect of the
particle cutoff in their lattice model by studying a deterministic
continuum front equation, in which the growth term was set to zero for
values of $\phi$  less than $1/N$. They showed that this led to a
correction to the asymptotic front speed of the order of $1/ \ln^2N$
with a prefactor, which is given in terms of the linear growth
properties of the equation without a cutoff. Because of the
logarithmic term, in the dominant order, it does not matter whether
the actual cutoff should really be exactly $1/N$ (corresponding to
exactly one particle), or whether the growth is just suppressed at
values of $\phi$ of order $1/N$, since $1/\ln^2(cN)$ $\approx$ $
1/\ln^2N$ in dominant order. Simulations of two different lattice
models by Brunet and Derrida \cite{bd} and by van Zon {\em et al.}
\cite{vanzon}  gave strong support for the essential correctness of
this  procedure for sufficiently large $N$, but showed that there can
be significant deviations from the asymptotic result for large but not
extremely large $N$. Moreover, for a different lattice model, Kessler
{\em et al.} \cite{kns} did observe a correction to the average front
speed of order $1/ \ln^2N$ but with a prefactor which they claimed was
a factor of order two different from the prediction of Brunet and
Derrida.

There are hence several questions that lead us to reconsider the
finite particle  effects on the average front speed of pulled
stochastic fronts:
\begin{itemize}
\item[{\em (i)}] Why is it that a simple cutoff of order $1/N$ in a
{\em deterministic equation for a continuum (mean-field type)
equation} apparently leads to the proper asymptotic correction to the
average speed of a {\em stochastic} front?
\item[{\em (ii)}] Can we get a more microscopic picture of the
stochastic behavior at the far end of the front, where there are only
a few particles per lattice site?
\item[{\em (iii)}] Can we go beyond the large $N$ asymptotic result of
Brunet and Derrida, e.g., can we  calculate the correction term for
large but not extremely large values of $N$ or even for arbitrary $N$?
--- After all, one might a priori expect correlation effects to be
very important for fronts whose propagation speed is strongly affected
by the region where there are only a few particles per site.
\item[{\em (iv)}] What is the role of correlation effects?
\item[{\em (v)}] To what extent do the specific details of the
particular stochastic model play a role?
\item[{\em (vi)}] Can one resolve the discrepancy noted by Kessler
{\em et al.} \cite{kns}?
\end{itemize}

\subsection{Summary of the main results}

In this paper, we address these questions and answer the majority of
them for a specific model for which Breuer {\em et al.} \cite{breuer}
already studied the asymptotic speeds of stochastic fronts numerically
a number of years ago. The model consists of particles making
diffusive hops on a one-dimensional lattice, and being subject to
growth and death on each lattice site. It is very close to the one
also studied by Kessler {\em et al.} \cite{kns}, the only difference
being that their model includes a correlation term, which is small and
irrelevant for large $N$. The absence of such correlations makes the
model studied by Breuer {\em et al.} easier to analyze. Moreover, an
examination of the numerical results therein shows that the deviation
of the asymptotic front speed from its pulled front value indeed
behaves as $1/\ln^2 N$, (although it was not realized in
\cite{breuer}), with a prefactor that, over the range of $N$-values
studied, is different from the one predicted later by Brunet and
Derrida \cite{bd}, but not as much different as Kessler {\em et al.}
claimed it to be for their own model \cite{kns}. For each stochastic
realization of a front, which moves into a region where no particles
are present, one can always identify a foremost occupied lattice site
or ``the foremost bin''. In the region near this one, fluctuations are
large and the discreteness of the lattice and of the particle number
occupation is extremely important: the standard description, which
assumes that the average particle density is uniformly translating,
breaks down in this region. Moreover, since the particle occupation
numbers are small in the tip, essentially all known methods fail,
based as they are on large-$N$ expansions.

With a finite particle cutoff, fronts are never really pulled, but
instead are weakly pushed \cite{deb2}. Even for the simplest case of a
pushed front in a second order nonlinear partial differential
equation, in general, the speed cannot be calculated explicitly. It
should therefore come as no surprise that with the various additional
complications described above,  we do   {\em not } have a full first
principles theory that gives the front speed for finite values of $N$
for the model we study. However, in this paper {\em we do obtain } a
number of new results for the behavior in the far tip of the front as
well as for the effect of the region behind the tip on the finite-$N$
corrections. These results can be tested independently and our
numerical simulations largely support the picture that emerges from
this approach.  In terms of short answers to the questions raised
above in Sec. \ref{IB}, we find that
\begin{itemize}
\item[{\em (i)}] For extremely large $N$, the asymptotic results of
Brunet and Derrida based on a simple cutoff of order $1/N$ in a {\em
deterministic equation for a continuum (mean-field type) equation}
become essentially correct because all the essential changes  are all
limited to a few bins behind the foremost one, where the particle
numbers are finite and small;  together with the fact that  $1/
\ln^2(cN) \approx 1/ \ln^2 (N) $ to dominant order, this ensures the
correctness of the asymptotic expression for $N\to \infty$.
\item[{\em (ii)}]  Yes, one can get a more microscopic picture of what
happens near the foremost bin of the front; we develop mean-field type
expressions for the probability distribution that describes the
``stop-and-go'' type behavior there (Sec. IV), and show that the
results compare well with numerical simulation results Sec. V.
\item[{\em (iii)}] A first-principles theory for the stochastic front
speed for arbitrary $N$ seems virtually impossible, except possibly in
some special limits, as in principle, it will involve matching the
approximately {\em uniformly translating} average profile behind the
tip of the front to the {\em non-uniformly translating}  near the
foremost bin, where standard methods do not seem to apply.
\item[{\em (iv)}] Correlation effects are very important near the tip;
we identify two of them and model one: rapid successive forward hops
of the foremost particle, Sec.~\ref{toofast} and jumping back of the
foremost particle, Sec.~\ref{backjump}.
\item[{\em (v)}] The details of the particular stochastic model play a
role for the corrections in the asymptotic front speed through the
global average front profile (quantified by $A$ of Secs. III and IV)
and through the effective profile near the tip, but their effects are
truly minute. We demonstrate this by means of a mean-field theory that
tries to extend the uniformly translating front solution all the way
to the far tip of the front (described in Sec.~\ref{seccorrec}). In
this theory, there is a quantity $a$ associated with the effective
profile at the tip, and we show that these two quantities, $A$ and
$a$, provide only subdominant corrections to the asymptotic large $N$
result.
\item[{\em (vi)}] The model considered by Kessler {\em et al.}
\cite{kns} is slightly different from the one considered by Breuer
{\em et al.}, in the sense that number of particles of each species is
finite. However, {\it a priori}, one expects that this difference in
the two models would not affect the speed corrections for large
$N$. Our own simulations confirm this, and give no sign of a
discrepancy between the asymptotic large-$N$ speed corrections
obtained from the two models (Sec.~\ref{knsmodel}).
\end{itemize}

We finally note that in this paper, we will focus on the case where
the growth and hopping terms for a few particles are the same as those
for a small but finite density of particles. In such cases, the front
speed converges for $N\to \infty$ to the pulled front speed $v^*$ of
the corresponding mean-field equation. As we will discuss elsewhere
\cite{PvS}, with only slight modifications of the stochastic rules for
few particles, one can also arrive at situations where the limits do
not commute, i.e., where the stochastic front speed converges to a
speed larger then $v^*$ as $N\to \infty$, even though the stochastic
model would converge to a  the mean-field equation with pulled fronts
in this limit.

\subsection{Complications Associated with Discreteness of the Lattice
and Particle Numbers}\label{complications}

The challenge of understanding the propagation of any one of these
fronts lies in the fact that as a consequence of the  discrete nature
of the particle events and of the particle number realizations, the
natural description of the far tip is {\em not in terms of a uniformly
translating solution for the average number of particles in the
bins\/} (we call each lattice site a ``bin''), but is in terms of {\em
discrete notions like the foremost bin, individual jumps, etc}. An
additional complication is that in the presence of fluctuations, the
front position exhibits diffusion-like wandering behaviour, which have
to be taken out in order to study the intrinsic stochastic front
dynamics, just like capillary waves beset analyzing the intrinsic
structure of a fluid interface (Sec.~\ref{wandering}).  The
implication of all this is that {\em (i)} in the presence of an
underlying lattice, instead of being uniformly translating, the
position of the foremost occupied bin advances in a discrete manner,
and {\em (ii) } due to the discrete nature of the constituent
particles, the position of the foremost bin advances
probabilistically, as its movement is controlled by diffusion.

Based on these ingredients and observations, the central theme in this
paper revolves around a picture of the tip of the front that is {\it
totally different from the conventional picture of a pulled front}. We
present the picture here in terms of its {\it simplified essence}, as
it is helpful for the reader to bear it in mind throughout this paper:
we call the foremost occupied lattice site at the far end of the tip
of the front ``the foremost bin''. Therefore, the very definition of
the {\em foremost bin } on a lattice site means that it is occupied by
at least one particle and that all the lattice sites on the right of
it are empty. Naturally, an empty lattice site (all the lattice sites
on the right of which are also empty) attains the status of the
foremost bin as soon as one  particle hops into it from the left. In
reference to the lattice, the position of the foremost bin remains
fixed at this site for some time, i.e., after its creation, a foremost
bin remains the foremost bin for some time. During this time, however,
the number of particles in and behind the foremost bin continues to
grow. As the number of particles grows in the foremost bin, the chance
of one of them making a diffusive hop on to the right also
increases. At some instant, a particle from the foremost bin hops over
to the right: as a result of this hop, the position of the foremost
bin advances by one unit on the lattice, or, viewed from another
angle, a new foremost bin is created which is one lattice distance
away on the right of the previous one. Microscopically, the selection
process for the length of the time span between two consecutive
foremost bin creations is stochastic, and the inverse of the long time
average of this time span defines the front speed. Simultaneously, the
amount of growth of particle numbers in and behind the foremost bin
itself depends on the time span between two consecutive foremost bin
creations (the longer the time span, the longer the amount of
growth). As a consequence, on  average, the selection mechanism for
the length of the time span between two consecutive foremost bin
creations, which determines the asymptotic front speed, is nonlinear.

This inherent nonlinearity makes the prediction of the asymptotic
front speed difficult. One might recall the difficulties associated
with the prediction of pushed fronts due to nonlinear terms in this
context, although the nature of the nonlinearities in these two cases
is {\it completely different\/}. In the case of pushed fronts, the
asymptotic front speed is determined by the mean-field dynamics of the
fronts, and the nonlinearties originate from the {\it nonlinear growth
terms in the partial differential equations\/} that describe the
mean-field dynamics  (As we discuss in Sec.~\ref{wandering},  if one
does not take out the wandering of the front positions, then the
nonlinear growth terms actually do affect the stochastic front
dynamics in a subtle way too).  On the other hand, for fronts
consisting of discrete particles on a discrete lattice, the
corresponding mean-field growth terms are {\it linear\/}, but since
the asymptotic front speed is determined from the probability
distribution of the time span between two consecutive foremost bin
creations, on average, it is the relation between this probability
distribution and the effect of the linear growth terms that the
nonlinearities stem from.

Our approach is to develop a separate probabilistic theory for the
hops to create the new foremost bins, and then to show that by
matching the description of the behaviour in this region to the more
standard one (of growth and roughly speaking, uniform translation)
behind it, one obtains a consistent and more complete description of
the stochastic and discreteness effects on the front propagation. In
the simplest approximation, the theory provides a very good fit to the
data, but our approach can be systematically improved by incorporating
the effect of fluctuations as well. Besides providing insight into how
a stochastic front propagates at the far tip of the leading edge, our
analysis naturally leads to a more complete description that allows
one to  interpret (though not predict) the finite $N$ corrections to
the front speed for much smaller values of $N$ than that are necessary
to see the asymptotic result of Brunet and Derrida \cite{bd}. As one
might expect, for values of $N$ where deviations from this asymptotic
result are important, model-specific effects do play a role.

For the major part of our analysis, we focus on the most relevant and
illuminating case in which {\em the diffusion and growth rates of the
model are both of the same order}. This regime is the most
illustrative as it displays all the aspects of finite particle and
lattice effects most clearly. We also investigate the case when the
diffusion rate is much smaller than the growth rate to illustrate the
correlation effects. For all of these cases, the matching between the
behaviour of the tip of the front and the standard description of a
uniformly translating solution behind it is a complicated process, for
the lack of a proper small parameter that allows one to do
perturbation theory.

The paper is organized in the following manner: in Sec. II, we
describe our model (which is the same as in Ref. \cite{breuer}) and
define the dynamics of the front. The crux of the paper is presented
in Sec. IV, where we present a detailed analysis of the microscopic
picture of the front dynamics and show that for the description of the
far tip of the front, one has to abandon the idea of a uniformly
translating front solution. The lattice and finite particle effects
lead to a ``stop-and-go'' type dynamics at the far tip of the front,
while the average front behind it ``crosses over'' to a uniformly
translating solution. In this formulation, the effect of stochasticity
on the asymptotic front speed is coded in the probability distribution
of the times required for the advancement of the foremost bin. We
derive expressions of these probability distributions by matching the
solution of the far tip with the uniformly translating solution
behind. This matching includes various correlation effects in a
mean-field type approximation. In Sec. V, we compare our theoretical
predictions of Sec. IV with the stochastic simulation results. In
addition to that, in Sec. III, we argue that the corresponding front
solution is a case of a weakly pushed front and analyze an effective
mean-field solution that extends all the way to the foremost bin
(thereby ignoring its breakdown near the foremost bin). This allows us
to rederive the asymptotic velocity expression of Brunet and Derrida
\cite{bd} and obtain the further subdominant finite-$N$ corrections to
it. In Sec. VI, we carry out the full stochastic simulation for the
model considered by Kessler {\it et al.\/}, and finally, we conclude
the paper with a discussion and outlook in Sec. VII.

\section{The Reaction-Diffusion Process X$+$Y 
$\rightleftharpoons$ 2X on a Lattice}

We consider the following reaction-diffusion process X$+$Y
$\rightleftharpoons$  2X on a  lattice in the following formulation:
at each lattice position, there exists a bin. We label the bins by
their serial indices $k$, $k=1,2,3,\ldots,M$, placed from left to
right. Each bin has an {\it infinite} supply of Y particles. An X
particle in the $k$-th bin can undergo three basic processes: {\em
(i)} diffusion to the ($k-1$)-th or the ($k+1$)-th bin with a rate of
diffusion $\gamma$. If an X particle in bin 1 jumps towards the left,
or an X particle in the $M$-th bin jumps to the right, then they are
immediately replaced, {\em (ii)} forward reaction to produce an {\it
extra} X particle having annihilated a Y particle (X $+$ Y
$\rightarrow$ 2X), with a rate $\gamma_g$, and {\em (iii)} if there
are at least two X particles present in the $k$-th bin, then any two
of the X particles can react with each other and annihilate one X
particle to produce a Y particle (2X $\rightarrow$ X $+$ Y), with a
reaction rate $\gamma_d$. A state of the system at time $t$ is given
by the numbers of X particles in the bins, denoted as
$\{N_1,N_2,\ldots,N_M;t\}$.

In the context of front propagation, the above model was first studied
by Breuer {\it et al\,} \cite{breuer}. Up to Sec. V of this paper, we
will confine ourselves to this model only. In Sec. VI, we will
consider a slightly modified version of this model, numerically
studied by Kessler and co-authors \cite{kns}, in which the number of Y
particles in any bin is finite, and the Y-particles can diffuse from
any bin to its nearest neighbour bins with the same diffusion rate
$\gamma$.

\subsection{The Master Equation}

The discrete, microscopic description of the above reaction-diffusion
process inherently introduces fluctuations in the number of X
particles present in any particular bin. This necessitates a suitable
multivariate probabilistic description of the system.  Let us denote
the probability of a certain configuration $\{N_1,N_2,\ldots,N_M;t\}$
at time $t$ by $P(N_1,N_2,\ldots,N_M;t)$.  The dynamics of
$P(N_1,N_2,\ldots,N_M;t)$ is given by the following master equation:
\bleq      \be   \frac{\partial}{\partial
t}\,P(N_1,\,N_2,\,\ldots,\,N_M;\,t)\,=\,\sum_k\{\gamma\,[\,(N_{k\,+\,1}\,+\,1)\,P(N_1,\,N_2,\,\ldots,\,N_k\,-\,1,\,N_{k\,+\,1}\,+\,1,\,\ldots,\,N_M;\,t)\nonumber\\&&\hspace{-9cm}+\,(N_{k\,-\,1}\,+\,1)\,P(N_1,\,N_2,\,\ldots,\,N_{k\,-\,1}\,+\,1,\,N_k\,-\,1,\,\ldots,\,N_M;\,t)\nonumber\\&&\hspace{-9cm}-\,2\,N_k\,P(N_1,\,N_2,\,\ldots,\,N_{k\,-\,1},\,N_k,\,N_{k\,+\,1},\,\ldots,\,N_M;\,t)\,]\nonumber\\&&\hspace{-9.8cm}+\,\gamma_g\,[\,(N_k\,-\,1)\,P(N_1,\,N_2,\,\ldots,\,N_{k\,-\,1},\,N_k\,-\,1,\,N_{k\,+\,1},\,\ldots,\,N_M;\,t)\nonumber\\&&\hspace{-8.7cm}-\,N_k\,P(N_1,\,N_2,\,\ldots,\,N_{k\,-\,1},\,N_k,\,N_{k\,+\,1},\,\ldots,\,N_M;\,t)\,]\,\nonumber\\&&\hspace{-9.8cm}+\,\frac{\gamma_d}{2}\,[\,N_k\,(N_k\,+\,1)\,P(N_1,\,N_2,\,\ldots,\,N_{k\,-\,1},\,N_k\,+\,1,\,N_{k\,+\,1},\,\ldots,\,N_M;\,t)\nonumber\\&&\hspace{-8.7cm}-\,N_k\,(N_k\,-\,1)\,P(N_1,\,N_2,\,\ldots,\,N_{k\,-\,1},\,N_k,\,N_{k\,+\,1},\,\ldots,\,N_M;\,t)\,]\,\}\,.
\label{e2.1}
\ee     \eleq
\noindent The above equation is actually not quite accurate at the 1st
and $M$-th boundary bins, but we refrain from writing out the
correction terms explicitly, as they are not needed in the analysis
below.

\subsection{The Macroscopic Density Field and the Fisher-Kolmogorov Equation}

If the forward reaction rate, $\gamma_g$, is much larger than the
annihilation rate $\gamma_d$, an initial conglomeration of X particles
will start to grow in size as well as in numbers. To study this growth
phenomena, we define $\langle N_k(t)\rangle$, the average number of X
particles in the $k$-th bin at time $t$, as     \be     \langle
N_k(t)\rangle=\!\!\!\!\!\sum_{\{N_{k'}\}_{k'=1\cdots
N}}\!\!\!\!\!N_k\,P(N_1,\,N_2,.\,..,\,N_M;\,t)\,.
\label{e2.2}
\ee     Using Eq. (\ref{e2.1}), it is easy to obtain the time dynamics
of $\langle N_k(t)\rangle$, given by      \be \frac{\partial}{\partial
t}\,\langle N_k(t)\rangle\,=\,\gamma\,\left[\,\langle
N_{k\,+\,1}(t)\rangle\,+\,\langle N_{k\,-\,1}(t)\rangle\,-\,2\,\langle
N_k(t)\rangle\,\right]\nonumber\\&&\hspace{-7.5cm}+\,\gamma_g\,\langle
N_k(t)\rangle\,-\,\frac{\gamma_d}{2}\,\left[\langle
N^2_k(t)\rangle\,-\,\langle N_k(t)\rangle\,\right]\,,
\label{e2.3}
\ee    with    \be    \langle
N^2_k(t)\rangle=\!\!\!\!\sum_{\{N_{k'}\}_{k'=1\cdots N}}
\!\!\!\!\!N^2_k\,P(N_1,\,N_2,.\,..,\,N_M;\,t)\,.
\label{e2.4}
\ee

For the sake of simplicity, we define $\tilde\gamma=\gamma/\gamma_g$,
$t'=\gamma_gt$ and $N=2\gamma_g/\gamma_d$, and reduce the number of
parameters in Eq. (\ref{e2.3}), to have \cite{breuer}   \be
\frac{\partial}{\partial t'}\langle
N_k(t')\rangle=\tilde\gamma\,\left[\,\langle
N_{k\,+\,1}(t')\rangle+\langle N_{k\,-\,1}(t')\rangle-2\,\langle
N_k(t')\rangle\,\right]\nonumber\\&&\hspace{-7.5cm}+\,\langle
N_k(t')\rangle\,-\,\frac{1}{N}\,\left[\langle
N^2_k(t')\rangle\,-\,\langle N_k(t')\rangle\,\right]\,.
\label{e2.5}
\ee   Following the procedure in Ref. \cite{breuer}, if one replaces
the $\displaystyle{\frac{1}{N}\left[\langle N^2_k(t)\rangle-\langle
N_k(t)\rangle\right]}$ term in Eq. (\ref{e2.5}) by
$\displaystyle{\frac{1}{N}\langle N_k(t)\rangle^2}$ and further
defines a mean ``concentration field'' on the $k$-th bin by
introducing the variable $\phi_k=\langle N_k\rangle/N$, then from
Eq. (\ref{e2.5}), one arrives at the following difference-differential
version of the Fisher-Kolmogorov equation for the reaction-diffusion
process X$+$Y $\rightleftharpoons$ 2X on a lattice, given by
\cite{breuer}   \be \frac{\partial}{\partial t}\,
\phi_k(t)\,=\,\tilde\gamma\,\left[\, \phi_{k\,+\,1}(t)\,+\,
\phi_{k\,-\,1}(t)\,-\,2\,
\phi_k(t)\,\right]\nonumber\\&&\hspace{-2.5cm}+\,
\phi_k(t)\,-\,\phi^2_k(t)\,.
\label{e2.6}
\ee

The original Fisher-Kolmogorov or F-KPP equation \cite{fisher,kpp} is
a partial differential equation in continuous space and time.  Notice
that in these variables, the properties of the propagating front
depend only on two parameters, $N$ and $\tilde\gamma$.

\section{Mean-Field approximations to the Propagating Front solution}

As mentioned earlier, in this section we do not consider the proper
stop-and-go type dynamics of the tip; instead, as a continuation of
mean-field equation (\ref{e2.6}) above, we describe the {\it
entire\/}\ front by the uniformly translating profile. We then make a
number of general observations concerning the uniformly translating
front solutions in mean-field type equations for the average profile,
from the perspective of the questions raised in the introduction. A
central result of the discussion will be an expression for the
finite-$N$ value of the velocity, which allows us to  interpret
deviations from the asymptotic results of \cite{bd} in terms  of
physical properties of stochastic fronts.

\subsection{Front propagation in the  dynamical equation for $\phi_k(t)$}

From the point of view of average number of X particles in the bins,
Eq. (\ref{e2.5}) has two stationary states. One of them, for which
$\langle N_k\rangle=N\quad\forall\,k$, is stable. The other, for which
$\langle N_k\rangle=0\quad\forall\,k$, is unstable. This means that
any perturbation around the unstable state  grows in time until it
saturates at the stable state value. In particular, if in a certain
configuration of the system, the stable and the unstable regions
coexist, i.e., $\langle N_k\rangle=N\quad\forall\,k<k_0$ and $\langle
N_k\rangle=0\quad\forall\,k>k_1$, with $k_1>k_0$, then the stable
region invades the unstable region and propagates into it. In other
words, in due course of time, the boundary between these two regions,
having a finite width, moves further and further inside the unstable
region. For a wide range of initial conditions, the speed, with which
this boundary moves into the unstable region, approaches a fixed
asymptotic value, $v_{as}$. Simultaneously, the shape of this boundary
between the two regions, determined by the average number of X
particles, $\langle N_k\rangle$, plotted against the corresponding bin
indices $k$, also approaches an asymptotic shape. This asymptotic
shape, therefore, becomes a function of $(k-v_{as}\,t)$ for long
times, and this well-known phenomenon is known as the front
propagation. In the present context, Eqs. (\ref{e2.5}-\ref{e2.6})
provides us with an example of front propagation into unstable
states. We will follow the usual convention that the front propagates
to the right in the  direction of increasing bin numbers.

In the mean-field approximation (\ref{e2.6}), the average particle
density field $\phi_k(t)$ obeys a difference-differential equation.
The asymptotic speed selection mechanism for propagating fronts into
unstable states has been a well-understood phenomenon for a number of
years, and it has been realized by various authors
\cite{bd,kns,levine,ebert} that the calculation of the  asymptotic
front speed on a lattice  for the type of Eqs. (\ref{e2.5}-\ref{e2.6})
proceeds along similar lines as it does for partial differential
equations.  It is well-known that for Eqs. (\ref{e2.5}-\ref{e2.6}),
the selection mechanism for $v_{as}$ depends entirely on the region,
where the nonlinear saturation terms $\displaystyle{\big(\left[\langle
N^2_k(t)\rangle-\langle N_k(t)\rangle\right]/N}$ or $\phi^2_k(t)\big)$
are much smaller in magnitude than the corresponding linear growth
terms [$\langle N_k(t)\rangle$ or $\phi_k(t)$], i.e., the leading edge
of the front, where the value of $\phi_k(t)$ is very small, such that
$\phi^2_k(t)\ll\phi_k(t)$. In this region, the nonlinear terms can be
neglected, and after having used
$\phi_k(t)\equiv\phi(k-v_{as}\,t)\equiv\phi(\xi)$, where
$\xi=k-v_{as}\,t\,$ is the comoving coordinate, Eq. (\ref{e2.6})
reduces to a linear difference-differential equation, given by \be
-\,v_{as}\,\frac{\partial}{\partial\xi}\,
\phi(\xi)\,=\,\tilde\gamma\,\left[\, \phi\,(\xi\,+\,1)\,+\,
\phi\,(\xi\,-\,1)\,-\,2\,\phi(\xi)\,\right]\,\nonumber\\&&\hspace{-2.5cm}+\,\phi(\xi)\,.
\label{e3.0}
\ee        If one {\it neglects\,} the fact that the microscopic X
particles are discrete and {\it assumes\,} that $\phi(\xi)$ goes to
zero continuously for $\xi\rightarrow\infty$, then a natural candidate
for the solution of $\phi(\xi)$ in the {\it linear\,}
difference-differential equation, Eq. (\ref{e3.0}) above, is
$\phi(\xi)\equiv\,A\,\exp[-z\xi]$, where $z$ is a {\it real\,} and
{\it positive} quantity.  With this solution of $\phi(\xi)$ in the
so-called leading edge of the front, one arrives at the dispersion
relation    \be
v_{as}\,\equiv\,v_{as}(z)\,=\,\frac{2\,\tilde\gamma\,[\,\cosh(z)\,-\,1\,]\,+\,1}{z}\,.
\label{e3.1}
\ee

Like the other examples of fronts propagating into unstable states,
Eq. (\ref{e3.1}) allows an uncountably infinite number of asymptotic
velocities depending on the selected value of the {\it continuous\,}
parameter $z$. However, for a steep enough initial condition that
decays faster than $\exp(-\lambda \xi)$ in $\xi$ for any $\lambda >z_0
$ determined below (hence, a unit step function obeys this condition),
the observed asymptotic speed  equals the so-called linear spreading
speed $v^*$, given by $v^*\equiv v^*(z_0)$, where $z_0$ is the
value of $z$, for which the dispersion relation $v_{as}(z)$ vs. $z$
has a minimum.

The fact that $v^*$ defined in this way is nothing but the linear
spreading speed, i.e., the spreading speed of small perturbations
whose dynamics is given by the linearized equation (\ref{e3.1}),
follows from a saddle point analysis of the asymptotic behavior of the
Green's function  for the linear equation (\ref{e3.1}), see
e.g. \cite{ebert}. The name pulled fronts stems from the fact that
this linear spreading almost literally ``pulls'' the nonlinear front
with it, the nonlinear terms just giving rise to saturation behind the
front.

\subsection{The deceptive subtlety of the mean-field
approximation}\label{wandering}

As we discussed above, in the pulled front regime, the front speed of
a given problem is determined completely by the linear term in the
dynamical equation. In going from the exact equation (\ref{e2.5}) to
the mean field approximation (\ref{e2.6}), we appear, at first sight,
to have ignored  only a term linear in $\phi_k$ of order $1/N$ [the
second term between  square brackets in (\ref{e2.5})]. Hence, naively
one might expect the front speed to converge as $1/N$ to the
asymptotic value $v^*(z_0)$. We already know from the work of Brunet
and Derrida \cite{bd}, however, that the convergence is much slower,
namely as $1/\ln^2N$. How can the two results be reconciled?

The resolution of the paradox lies in the fact that in the mean field
approximation we completely ignore the diffusive wandering of
fronts. If we follow the evolution of an ensemble of fronts, their
positions [defined, e.g., by Eq. ~(\ref{e4.1}) below] will fluctuate:
the root mean square wandering of the fronts grows as $\sqrt{t}$ as
for any one-dimensional random walker \cite{breuer,armero}. This means
that {\em in reality the ensemble average $\langle N_k(t)\rangle $
does not acquire a fixed  shape in the frame moving with the average
speed. Instead, the average profile $\langle N_k(t)\rangle$ continues
to broaden in time, although the front shapes for the individual
realizations reach an asymptotic shape\/}\ (see Fig. 5 of
Ref. \cite{breuer} for an illustration). This has a severe
consequence: we cannot simply assume that the $\langle
N^2_k(t)\rangle$ term is small in the leading edge of the profile
where $\langle N_k(t)\rangle$ is small, and replace it by $\langle
N_k(t)\rangle^2$ --- the few members of the ensemble, which are
relatively further ahead, do give significant contributions through
this term in regions where $\langle N_k(t)\rangle$ is small. Thus,
while Eq. (\ref{e2.5}) is exact and contains the fluctuation effects
due to the root mean square wandering of the front, the mean-field
approximation (\ref{e2.6}) throws out such effects completely.

If, on the other hand, we look at the shape of a particular front
realization in the appropriate position, so that the front wandering
is taken out, the mean-field equation does yield a reasonably good
description of this (conditionally averaged) front profile in the
range where the particle occupation numbers are large and (hence)
where fluctuation effects are small. Additional information is needed,
however, to calculate the front speed.

In passing, we note that the situation is somewhat similar to the
theory of fluid interfaces: capillary wave fluctuations wash out the
average interface profile completely, but on scales of the order of
the capillary length, the mean field theory for the so-called
intrinsic interface profile works quite well.

\subsection{The front speed correction for large $N$}\label{seccorrec}

The above observations already allow us to arrive at and extend the
results of Brunet and Derrida \cite{bd} from a slightly different
angle than in their original work as follows. First of all, from the
discussion above, we notice that even though a mean field
approximation (\ref{e2.6}) does not work for the ensemble-averaged
front profile, but for a given stochastic front realization, the mean
field theory does apply to a good approximation in the bins, where the
number of X particles are relatively large. These are essentially the
bins that are sufficiently behind the {\em foremost bin}, the
rightmost bin in given stochastic realization, on the right of which
all bins are completely empty. Nevertheless, as mentioned in the
beginning paragraph of this section, we assume that the uniformly
translating front solution of Eq. (\ref{e2.6}) holds for the
description of the front profile all the way up to the foremost bin
for a given realization. Secondly, the actual front solution of
Eq. (\ref{e2.6}) is a case of a {\it weakly pushed\/} front as opposed
to being a truly pulled front \cite{deb2,notepushed}. This can be
understood in the following manner: notice that in any bin the forward
reaction X $+$ Y $\rightarrow$ 2X does not proceed unless there is at
least one X particle in that bin to start with. As for any given
realization of the stochastic front, the front propagation on a
lattice is tantamount to the discrete forward movement of the foremost
bin by units of 1 (which can happen only through the diffusion of an X
particle from the foremost bin towards the right), in the uniformly
translating front solution of Eq. (\ref{e2.6}), the dynamics of the
tip of the front is diffusion dominated. This makes any given
realization of the front weakly pushed as opposed to being truly
pulled, and moreover, the asymptotic speed $v_N$ is expected to be
$<v^*$ for a finite $N$. This indicates that if we want to build all
these in the same frame as in the velocity selection mechanism for a
pulled front, one has to allow complex values of the parameter $z$
(see Eq. (\ref{e3.1}) and the discussion thereabove). Furthermore, the
existence of a foremost bin requires that the front profile must have
a zero {\it a bin ahead of the foremost bin}. Having combined all
these together, and without any loss of generality, we now require
that the front profile in the linear region of Eq. (\ref{e2.6}) is
given by \cite{bd,kns,levine,vs2,ebert} for  $\phi(\xi)$ for $v_N <
v^*$ \be
\phi(\xi)\,=\,A\,\sin\,[\,z_i\,\xi\,+\,\beta\,]\,\exp(-\,z_r\,\xi)\,,
\label{e3.3}
\ee such that $\phi(\xi)$ has a node at the coordinate of the bin just
ahead of the foremost bin. In Appendix \ref{bdextension}, we show how
Eq. (\ref{e3.3}) can be used to determine the complex decay rate $z$
in terms of $N$ and other parameters, and from that we obtain the
deviation of the front speed $v_N$ from $v^*$. The front speed $v_N$
is given by \bleq \be
v_N\,=\,v^*\,-\,\frac{d^2v_{as}}{dz^2}\bigg|_{z_0}\!\!z^2_i\,+\,O(z^4_i)\,\,\approx\,\,v^*\,-\,\frac{d^2v_{as}}{dz^2}\bigg|_{z_0}\,\frac{\pi^2\,z^2_0}{\left[\,\ln
N+z_0+\displaystyle{\ln\frac{A}{a}+\ln\left\{\sin\frac{\pi\,z_0}{\ln
N+1}\,\right\}}\right]^2}\, ,
\label{e3.12}
\ee       \eleq
\noindent    where, according to Eq. (\ref{e3.1}), \be
\frac{d^2v_{as}}{dz^2}\bigg|_{z_0}\,=\,\frac{\tilde\gamma\,\cosh\,z_0}{z_0}\,.
\label{e3.13}
\ee In the limit of large $N$,  the above result (\ref{e3.12}) reduces
to \be 
v_N\,\approx\,\,v^*\,-\,\frac{d^2v_{as}}{dz^2}\bigg|_{z_0}\,\frac{\pi^2\,z^2_0}{\ln^2N}\,
,
\label{e3.14}
\ee which is nothing but the asymptotic expression for the velocity
correction derived by Brunet and Derrida \cite{bd}.  Their approach is
based on  the  {\it partial differential equation\/} analog of the mean
field dynamical equation (\ref{e2.6}), in they introduced  an
artificial cutoff for the growth term for values of
$\phi(\xi)<\varepsilon$, where $\varepsilon\approx1/N$, to mimic the
dominant role played by diffusion at the tip of the front as opposed
to the growth term.

\subsection{Implications and discussion}

The above expressions for the speed corrections are already quite
instructive. First of all, as we pointed out, for the speed difference
$v^*-v_N$, Eq.~(\ref{e3.13}) reduces to the expression of
Eq.~(\ref{e3.14}) of Brunet and Derrida \cite{bd} at the dominant
order in the limit of very large $N$. To this order, the speed change
is given {\em explicitly} in terms of $N$. The more general
expression, Eq.~(\ref{e3.13}), however, contains the factors $A$ and
$a$; these affect the {\em sub}dominant behaviour, i.e., the
corrections to the asymptotic large $N$ expression.  For realistic
values of $N$, the corrections to the asymptotic behavior can be quite
significant  \cite{bd}.  As we shall show in Sec.~\ref{Asection}, $A$
depends on the global behaviour of the average front solution,
including the behaviour in the region where nonlinearities are
important. {\it This makes its value vary from model to model and it
is at this place where the specific details of the model affect the
speed difference $v^*-v_N$\/}. On the other hand, $a$ is only a
parameter that originates through the extrapolation of the mean-field
profile (\ref{e3.3}) to the foremost bin region. We will show in the
next section that the quantity $a$ is a fictitious quantity, as the
average front profile deviates significantly from the one in
Eq. (\ref{e3.3}) near the foremost bin: as we shall see, unlike the
mean-field solution,  it is not even uniformly translating.  {\em This
is the reason that an explicit general prediction  for the  fronts
speed beyond the asymptotic result obtained by Brunet and Derrida
\cite{bd} is hard, if not impossible, to come by}.

In passing, we note the following. It is well known from the analysis
of uniformly translating front solutions of the Fisher-Kolmogorov
partial differential equation that front solutions with $v<v^*$ are
asymptotically given by an expression like (\ref{e3.3}), and that
these fronts solutions with nodes  are unstable. This does not mean,
however, that the above (crude) analysis is  based on an unstable
solution  (\ref{e3.3}) and therefore inconsistent. The point is that
the expression (\ref{e3.3}) is only an intermediate asymptotic
solution, valid over some finite range of bins;  just as  in the
analysis of the slow time relaxation of pulled fronts in partial
differential equations \cite{ebert}, where such solutions also play a
role as intermediate asymptotics, but they do not make the full
solution unstable.
 
\section{The Probabilistic Dynamics of the Tip: Breakdown of the
Definition of the Comoving coordinate $\xi$}

We now turn to the analysis of the stochastic dynamics near the
foremost bin, which is the region which determines most of the front
dynamics. In the light of the discussion of Sec. \ref{wandering}, from
here onwards, we confine ourselves to the study of {\it
one single front realization\/}. 

Let us assume that as the front moves in time from the left to the
right, at some time $t=t_0$, the bin $k_0$ is deep inside the
saturation phase of the front. At time $t\geq t_0$, the total number of
particles on the right of the $k_0$-th bin is given by   \be
N_{tot}(t)\,=\,\sum_{k>k_0} N_k(t)\,.
\label{e4.1}
\ee    For large $t-t_0$, $N_{tot}(t)$ grows linearly and one may
define the asymptotic front speed $v_N$ as    \be
v_N\,=\,\frac{1}{N}\lim_{t\rightarrow\infty}\,\frac{N_{tot}(t)\,-\,N_{tot}(t_0)}{t\,-\,t_0}\,.
\label{e4.2}
\ee    Simultaneously, the position of the foremost bin also shifts
towards the right. For long times, the average rate at which the
position of the foremost bin shifts towards the right is the same as
the front speed measured according to the definition Eq.~(\ref{e4.2}),
as otherwise, an individual front realization will never reach an
asymptotic shape. 

\begin{figure}[h]
\begin{center}
\includegraphics[width=0.48\textwidth]{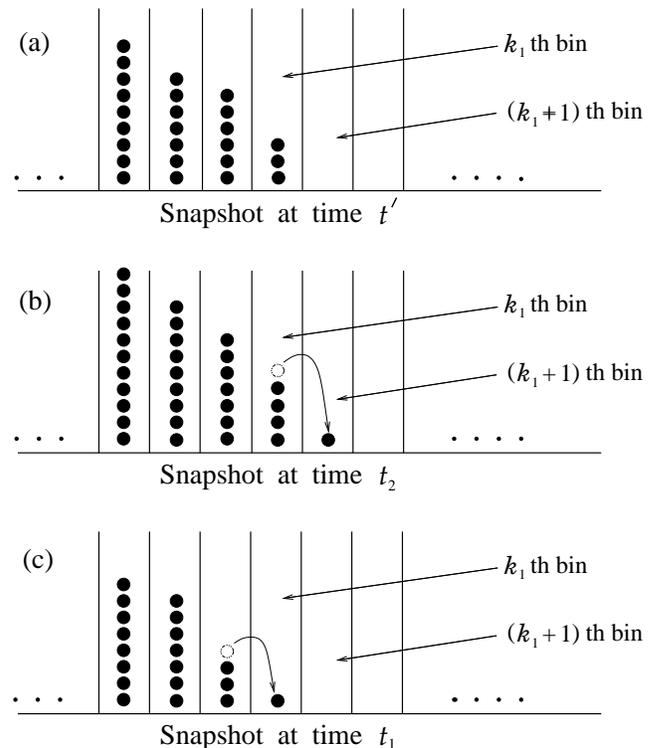}
\caption{Snapshots of one particular realization at times $t'$, $t_2$
and $t_1$. The filled circles denote the X particles in different
bins. At time $t_2$, $(k_1+1)$-th bin becomes the new foremost bin. In
a similar manner, $k_1$-th bin became the new foremost bin at time
$t_1$.\label{fig1}}
\end{center}
\end{figure}

Let us now examine the dynamics of the foremost bin in one particular
realization. In Sec.~III. The foremost bin moves towards the right by
means  of hops of the X particles. The way this diffusion takes place
is as follows: let us imagine that in one particular realization, at a
certain time $t'$, the index for the foremost bin is $k_1$, i.e., at
time $t'$, all the bins on the right of the $k_1$-th bin in that
realization are  not occupied by the X particles (see
Fig. \ref{fig1}(a)). The diffusion of the X particles from the
$k_1$-th bin to the $(k_1+1)$-th bin is not a continuous process. As a
result, it takes some more time before the first X particle diffuses
from the $k_1$-th bin to the $(k_1+1)$-th bin. Let us denote, by
$t_2$, the time instant at which this diffusion takes place (see
Fig. \ref{fig1}(b)). Clearly, there is no exchange of X particles
between the $k_1$-th bin and the $(k_1+1)$-th bin in the time interval
$t'\leq t<t_2$. During this time however, there can be time spans,
where the number of the X particles in the $k_1$-th bin may drop down
to zero, since in the time interval $t'\leq t< t_2$, the diffusion of
the X particles out of the $k_1$-th bin towards its left is an allowed
process. By definition, at time $t_2$, the $(k_1+1)$-th bin becomes
the ``new foremost bin''. Let us now denote, by $t_1$, the time
instant when the $k_1$-th bin became the ``new foremost bin'' due to
the diffusion of an X particle from the $(k_1-1)$-th bin in exactly
the same manner (see Fig. \ref{fig1}(c)). In this notation, therefore,
$t_2>t_1$, and we say that $k_1$-th bin remains the foremost bin for
the time interval $\Delta t=t_2-t_1$. If we now have a series of such
$\Delta t$ values {\it in sequence\/}, i.e., a sequence of time values
$\Delta t_1,\Delta t_2,\ldots,\Delta t_j$, for which a bin remains the
foremost bin, then it is easily seen that the asymptotic front speed
is also given by    \be
v_N\,=\,\lim_{j\rightarrow\infty}\,j\,\left[\sum_{j'=1}^{j}\Delta
t_{j'}\right]^{-1}\,.
\label{e4.3}
\ee Put in a different way, if we denote the probability that a
foremost bin remains the foremost bin for time $\Delta t$ by ${\cal
P}(\Delta t)$, the asymptotic front speed, according to
Eq. (\ref{e4.3}), is given by \be v_N\,=\,\left[\,\int_0^\infty
d(\Delta t)\,\,\Delta t\,\,{\cal P}(\Delta t)\,\right]^{-1}\,.
\label{e4.4}
\ee
Henceforth, our goal is to obtain a theoretical expression for ${\cal
P}(\Delta t)$, for given parameter values $N$ and $\tilde\gamma$.  As
a first approach, we will make an attempt to devise a mean-field
theory for this purpose. It is precisely at this place that we need to
study the origin and the consequences of the breakdown of the
definition of the comoving coordinate, $\xi$.

\subsection{The Stalling Phenomenon: lowest order approach} \label{sec4a}

The origin of the breakdown of the definition of the comoving
coordinate, $\xi$, in a mean field description is quite easy to
understand. As can be seen from the discussion in the paragraph above
Eq. (\ref{e4.3}), the key lies in the fact that for the time
a foremost bin remains the foremost bin, the front in the tip
region does not move at all. We refer to this  as the ``stalling
phenomenon''. During such stalling periods, all the dynamics is
confined within the left of (including) the foremost bin.  It is this
stalling phenomenon that is responsible for the breakdown of the
definition of the comoving coordinate, $\xi$ \cite{stallingnote}.
\vspace*{0.4cm}

\begin{tabular}{||p{0.7cm}|p{6.8cm}||}
\hline \hline $k_f$ & the label of the foremost bin between time $t=0$
and $t=\Delta t$ in an actual realization, e.g., in a computer
simulation.\\ \hline $k_m$ & the label of the bin that attains the
status of the foremost bin at time $t=0$ in the mean field theory that
we describe in this section. Naturally, at $t=0$, the density of X
particles in it is equal to $1/N$. \\ \hline $k_{m_0} $ &   the label
of the bin, where the  average front profile $\phi^{(0)}$,
extrapolated from behind, is equal to $1/N$. \\ \hline  $k_b$ &   the
label of the bin behind the tip, from which point on corrections to
the profile $\phi^{(0)} $ are neglected.\\ \hline $k_n$ & The bin
where $\phi^{(0)}$ becomes zero, i.e., the value of $k$ where the
argument of the $\sin$ function of $\phi^{(0)}$ becomes $\pi$.\\
\hline \hline
\end{tabular}

{\footnotesize Table I: Summary of the various coordinate labels used
  in the paper.}
\vspace*{0.4cm}

Our first step in analyzing the stalling phenomenon is to get back to
the $k$ and the $t$ coordinates, but in a different way than  we have
used them so far: the foremost bin, for the entire duration it remains
the foremost bin, is indexed by {\it an arbitrary fixed integer\/}
$k_f$ in this new scheme of relabelling the bin indices. The rest of
the bins are accordingly indexed by their positions with respect to
the $k_f$-th bin. Moreover, we start to count time (i.e., set the
clock at $t=0$) as soon as an X particle diffuses into the $k_f$-th
bin from the left and stop the clock just when an X particle diffuses
from the $k_f$-th bin to the right. This relabelling strongly
resembles the system of comoving coordinates, hence we call it the
``quasi-comoving coordinates''. In this formulation, {\it the clock
stops at time $\Delta t$ and resets itself to zero\/}. In this manner,
{\it the propagation of the front is a repetitive process of creating
new foremost bins in intervals of $\Delta t$\/}. Of course, it is a
probabilistic process, in which the value of $\Delta t$ is not fixed.

Our mean-field theory essentially mimics the stalling phenomenon just
as we see it in a computer simulation. In this theory, we also have a
foremost bin, which we index by a fixed integer $k_m$ in the
quasi-comoving frame. In these coordinates, we describe the dynamics
of the front by the average number of X particles in the bins. Between
the times $t=0$ and $t=\Delta t$, all the dynamics of the front is
confined to the left of (including) the $k_m$-th bin. For the benefit
of the reader, we summarize  the various coordinates $k$ used in this
paper in Table I.

The equations of motion in this quasi-comoving frame, analogous to
Eq. (\ref{e2.5}), in terms of the bin indices $k$ are therefore given
by     \be       \frac{\partial}{\partial t}\langle
N_k(t)\rangle=\tilde\gamma\,\left[\,\langle
N_{k\,+\,1}(t)\rangle+\langle N_{k\,-\,1}(t)\rangle-2\,\langle
N_k(t)\rangle\,\right]\nonumber\\&&\hspace{-8cm}+\,\langle
N_k(t)\rangle\,-\,\frac{1}{N}\,\left[\langle
N^2_k(t)\rangle\,-\,\langle
N_k(t)\rangle\,\right]\,\quad\forall\,k<k_m\,,
\nonumber\\&&\hspace{-8cm}\frac{\partial}{\partial t}\langle
N_k(t)\rangle=\tilde\gamma\,\left[\,\langle
N_{k\,-\,1}(t)\rangle-\,\langle N_k(t)\rangle\,\right]\,+\,\langle
N_k(t)\rangle\,\nonumber\\&&\hspace{-6.6cm}-\,\frac{1}{N}\,\left[\langle
N^2_k(t)\rangle\,-\,\langle
N_k(t)\rangle\,\right]\,\quad\mbox{for}\,\,k=k_m\,,\nonumber\\&&\hspace{-7cm}\mbox{and}\,\,\,
\langle N_k\rangle \,=\,0\,\quad\quad\forall\,k>k_m\,,
\label{e4.5}
\ee     for $0<t<\Delta t$, with the initial condition that $\langle
N_{k_m}\rangle=N_{k_m}=1$ at time $t=0$. The angular brackets above
denote quantities averaged over many snapshots of one single front
realization at time $t$. We focus our attention to the region at the
leading edge of the front (up to the $k_m$-th bin), where the
nonlinearities can be neglected so that the dynamics is given by
\be   \frac{\partial}{\partial t}
\,\phi_k(t)=\tilde\gamma\,\left[\,\phi_{k\,+\,1}(t)+
\phi_{k\,-\,1}(t)-2\, \phi_k(t)\,\right]\,+\,
\phi_k(t)\,,\nonumber\\&&\hspace{-3cm}\quad\forall\,k<k_m
\nonumber\\&&\hspace{-7.7cm}\frac{\partial}{\partial t}
\,\phi_k(t)=\tilde\gamma\,\left[\, \phi_{k\,-\,1}(t)-\,
\phi_k(t)\,\right]\,+\, \phi_k(t)\,,\nonumber\\&&\hspace{-3cm}
\quad\mbox{for}\,\,k=k_m
\label{e4.6}
\ee    with $\phi_k(t)=\langle N_k(t)\rangle/N$, $0<t<\Delta t$ and
$\phi_{k_m}=1/N$ at time $t=0$. Equation (\ref{e4.5}) explicitly
illustrates that the growth of the probability ahead of the foremost
bin is somewhat different from that behind the foremost bin as a
result of the stalling.

Before, we already introduced the probability ${\cal P}(\Delta t)$
that the foremost bin remains the foremost one between the times $t=0$
and $t=\Delta t$. Since the foremost bin ceases to be the foremost one
when a particle jumps out of it to the neighboring empty one on the
right, ${\cal P}(t)$ obeys the differential equation      \be
\frac{d}{dt}\,{\cal P}(t)\,=\,-\,\tilde\gamma\,\langle
N_{k_m}(t)\rangle\,{\cal P}(t)\,,
\label{e4.7}
\ee     or equivalently,     \be     {\cal P}(\Delta
t)=\tilde\gamma\,\langle N_{k_m}(\Delta
t)\rangle\,\exp\!\left[-\,\tilde\gamma\,\int_0^{\Delta t}\!dt\,\langle
N_{k_m}(t)\rangle\right],
\label{e4.8}
\ee     satisfying the normalization condition. Clearly, as one can
see from Eqs. (\ref{e4.4}) and (\ref{e4.8}), the proper asymptotic
speed is determined by $\langle N_{k_m}(t)\rangle$, which in turn must
come out of the solution of Eq. (\ref{e4.6}), i.e., from the effect of
the stalling phenomenon on the leading edge of the front.

The dynamics of the leading edge of the front, described by our
mean-field theory in the previous two paragraphs, is a  clear
over-simplification. In an actual realization, the dynamics of the tip
that  governs the probability distribution ${\cal P}(\Delta t)$ in the
quasi-comoving frame, is quite complicated. The foremost bin has only
a few particles, and as a consequence, the fluctuation in the number
of particles in it plays a very significant role in deciding the
nature of the probability distribution ${\cal P}(\Delta t)$. Arising
out of the fluctuations, there are two noteworthy events that have
serious consequences for the behaviour of ${\cal P}(\Delta t)$: {\em
(i)} The creation of the new foremost bins is a probabilistic process,
for which the time scale is characterized by $1/v_N$. However, if
several foremost bins are created {\it in a sequence\/} relatively
fast compared to the time scale set by $1/v_N$, then one naturally
expects that soon there would be a case when the new foremost bin
would be created at an unusually large value of $\Delta t$. {\em (ii)}
According to our definition, in the actual realization of the system,
the $k_f$-th bin remains the foremost bin between time $t=0$ and
$t=\Delta t$. However, it may so happen that during this time, all the
X particles in the $k_f$-th bin diffuse back to the left, leaving it
empty for some time, until some other X particle hops into it, making
it non-empty back again at a time $0<t<\Delta t$. By the nature of
construction, no mean-field theory can ever hope to capture the
fullest extent of these fluctuations, and the one that we just
presented above [that represents the effect of the stalling phenomenon
on the asymptotic speed selection mechanism for the front by
considering ${\cal P}(\Delta t)$], is no exception. Therefore, in this
mean field theory that we described in this section, such fluctuation
effects are completely suppressed. We will return to these fluctuation
effects in Sec. \ref{sec4c} below, where we will make an attempt to
estimate the effects of  these fluctuations on ${\cal P}(\Delta
t)$. The corresponding estimates will then be used to improve the
theoretical prediction of ${\cal P}(\Delta t)$ as well as to draw
limits on the validity of our mean-field theory.

\subsection{Effect of the Stalling Phenomenon on the Front Shape near
the Foremost Bin \label{stalling}}

In the previous subsection, we obtained a mean field type expression
for ${\cal P}(\Delta t)$ in terms of $\langle N_{k_m}(t)\rangle$. A
first approximation for $\langle N_{k_m}(t)\rangle$ would be obtained
from the solution of Eq. (\ref{e4.5}) above. However, in practice the
average occupation $\langle N_{k_m}(t)\rangle$ is affected by the
stalling effect itself. We now account for this effect in a
self-consistent way by calculating the corrections to the front shape
near the foremost bin. We start with Eqs. (\ref{e4.6}-\ref{e4.7}), and
subsequently build upon the considerations of Sec. III, where we
derived the solution $\phi(\xi)=A\sin[z_i\xi]\exp(-z_r\xi)$ at the
leading edge of the front.

\begin{figure}[h]
\begin{center}
\includegraphics[width=0.48\textwidth]{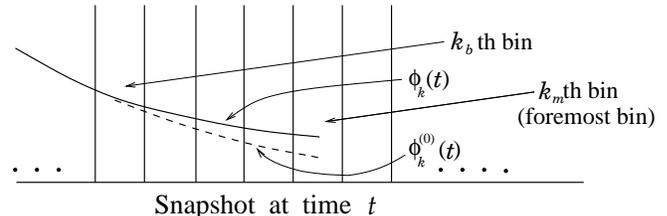}
\caption{Snapshot of the tip of the front in a mean-field description
at time $0<t<\Delta t$, showing density buildup of X particles on and
behind the foremost bin for large enough value of $t$. The dotted
curve is for the ``normal solution'',
$\phi^{(0)}_k(t)=A\sin[z_i(k-v_Nt)+\beta]\exp[-z_r(k-v_Nt)]$. The
solid curve is for the actual function $\phi_k(t)$. Even though both
$\phi^{(0)}_k(t)$ and $\phi_k(t)$ are discrete functions of $k$, we
have drawn continuous curves for clarity. \label{fig2}}
\end{center}
\end{figure}

A naive approach would be to claim that the shape of the leading edge
of the front, described by the set of equations (\ref{e4.6}), is given
by $\phi_k(t)=A\sin[z_i(k-v_Nt)+\beta]\exp[-z_r(k-v_Nt)]$ for
$0<t<\Delta t$ in the quasi-comoving frame. Notice that we have
reintroduced the phase factor $\beta$ inside the argument of the sine
function, in view of the fact that $k$ can only take integral values.
This solution of $\phi_k(t)$ would once again generate the same
dispersion relation as in Eq. (\ref{e3.4}). However, it is intuitively
quite clear that this solution of $\phi_k(t)$ cannot hold all the way
upto $k=k_m$, since the equations of motion for $k<k_m$ are different
from the equation of motion for $k=k_m$. First of all,
$\phi_{k_m}(t=0)=1/N$, which may not necessarily be equal to the value
of the function $A\sin[z_i(k_m-v_Nt)+\beta]\exp[-z_r(k_m-v_Nt)]$ at
time $t=0$. Secondly, for the entire duration of $0<t<\Delta t$, the
tip of the front is stationary at $k_m$, and as a result, the flow of
particles from the left starts to accumulate in the $k_m$-th
(foremost) bin. With increasing value of $t$, bins on the left of the
foremost bin get to know that the tip of the front has stalled, and
the correlation among different bins starts to develop on the left of
the foremost bin. As a result, an excess of particle density beyond
the corresponding ``normal solution'' values
$A\sin[z_i(k-v_Nt)+\beta]\exp[-z_r(k-v_Nt)]$ builds up on the left of
(including) the foremost bin over time. This is demonstrated in
Fig. \ref{fig2}.

To deal with the effect of stalling phenomenon on the density of X
particles in the bins at the tip of the front, which is very crucial
to calculate $\langle N_{k_m}(t)\rangle$, let us express $\phi_k(t)$
as  \be   \phi_k(t)\,=\,\phi^{(0)}_k(t)\,+\,\delta\phi_k(t)\,,
\label{e4.9}
\ee   where
$\phi^{(0)}_k(t)=A\sin[z_i(k-v_Nt)+\beta]\exp[-z_r(k-v_Nt)]$.  The
quantity $\delta\phi_k(t)$ then denotes the deviation of the density
of the X particles in the $k$-th bin from the ``normal solution''
$\phi^{(0)}_k(t)$. It takes time for the deviation to develop in any
bin, and moreover, since such correlation effects spread diffusively,
the information that the tip of the front has stalled at the foremost
bin does not affect too many bins behind the foremost bin.  Thus, it
is reasonable to assume that on the left of the foremost bin, there
exists a bin, henceforth indexed by $k_b$ in this quasi-comoving
coordinate (i.e., $k_b<k_m$), where the magnitude of $\delta\phi_k(t)$
is so small that we can impose the condition that
$\delta\phi_{k_b}(t)=0$. We then substitute Eq. (\ref{e4.9}) in
Eq. (\ref{e4.6}) and without having to worry about the equation of
motion for $\delta\phi_{k_b}(t)$, we obtain the equations of motion of
the quantities $\delta\phi_k(t)$ for $k_b<k\leq k_m$ as   \bleq    \be
\frac{\partial}{\partial t}
\,\delta\phi_k(t)=\tilde\gamma\,\left[\,\delta\phi_{k+1}(t)-2\,
\delta\phi_k(t)\,\right]\,+\,
\delta\phi_k(t)\,,\quad\quad\mbox{for}\,\,k\,=\,k_b\,+\,1\,,\nonumber\\&&\hspace{-10.5cm}
\frac{\partial}{\partial t}
\,\delta\phi_k(t)=\tilde\gamma\,\left[\,\delta\phi_{k\,+\,1}(t)+
\delta\phi_{k\,-\,1}(t)-2\, \delta\phi_k(t)\,\right]\,+\,
\delta\phi_k(t)\,\quad\quad\forall\,\,(k_b\,+\,1)<k<k_m\nonumber\\&&\hspace{-10.5cm}\frac{\partial}{\partial
t} \,\delta\phi_k(t)=\tilde\gamma\,\left[\,\delta\phi_{k\,-\,1}(t)-\,
\delta\phi_k(t)\,\right]\,+\,
\delta\phi_k(t)\,-\,\tilde\gamma\,[\,\phi^{(0)}_{k+1}\,-\,\phi^{(0)}_k\,]\,\quad\quad\mbox{for}\,\,k\,=\,k_m
\label{e4.10}
\ee    If we now denote the $(k_m-k_b)$-dimensional column vector
$[\,\delta\phi_{k_m}(t),\,\delta\phi_{k_m-1}(t),\ldots,\delta\phi_{k_b+1}(t)\,]$
by $\delta\Phi(t)$, then Eq. (\ref{e4.10}) becomes an inhomogeneous
linear differential equation in $\delta\Phi(t)$, given by    \be
\frac{d}{dt}\,\delta\Phi(t)\,=\,{\mathbf
M}\,\delta\Phi(t)\,+\,\delta\Phi_p\,,
\label{e4.11}
\ee    where ${\mathbf M}$ is the
$(k_m-k_b)\times(k_m-k_b)$-dimensional tridiagonal  symmetric matrix:
\be  {\mathbf
M}\,=\,\left[\begin{array}{cccccc}{1-\tilde\gamma}&\tilde\gamma&0&\ldots&0&0\\\tilde\gamma&{1-2\tilde\gamma}&\tilde\gamma&0&\ldots&0\\0&\tilde\gamma&{1-2\tilde\gamma}&\tilde\gamma&\ldots&0\\.&.&.&.&.&.\\0&\ldots&0&\tilde\gamma&1-2\tilde\gamma&\tilde\gamma\\0&\ldots&0&0&\tilde\gamma&{1-2\tilde\gamma}\end{array}\right]\,,
\label{e4.12}
\ee    and
$\delta\Phi_p=[\,\tilde\gamma\,(\phi^{(0)}_{k_m}-\phi^{(0)}_{k_m+1}),\,0,\ldots,0\,]$.
The solution of the linear inhomogeneous differential equation,
Eq. (\ref{e4.11}), is straightforwardly obtained as    \be
\delta\Phi(t)\,=\,\exp[{\mathbf
M}t]\,\,\delta\Phi(t=0)\,+\,\int_0^{t}dt'\,\exp[{\mathbf
M}(t-t')]\,\,\delta\Phi_p(t')\,.
\label{e4.13}
\ee    \eleq

To obtain the expression of $\langle N_{k_m}(t)\rangle$, which is our
final goal, we have to determine the unknowns $\delta\Phi(t=0)$. Of
these, the expression of $\delta\phi_{k_m}(t=0)$ is already known from
the fact that at time $t=0$, there is exactly one X particle in the
$k_m$-th bin, i.e.,    \be
\delta\phi_{k_m}(t=0)\,=\,\frac{1}{N}\,-\,\phi^{(0)}_{k_m}(t=0) \,.
\label{e4.14}
\ee      The values of $\delta\phi_{k}(t=0)$ for $k_b<k<k_m$ are also
quite easily determined when we notice that at time $t=\Delta t$, the
values of $\delta\phi_{k}(t=\Delta t)$ must reach the corresponding
values of $\delta\phi_{k-1}(t=0)$, because the average shape of the
front repeats itself once every $\Delta t$ time (note here that the
repetitive character of foremost bin creation in the quasi-comoving
frame is built in).  This leads us to the following set of $k_m-k_b-1$
consistency conditions \be  \delta\phi_{k_b+1}(t=0)\,=\,\int_0^\infty
d(\Delta t)\,{\cal P}(\Delta t)\,\delta\phi_{k_b+2}(\Delta
t)\nonumber\\&&\hspace{-4.6cm}\vdots\nonumber\\&&\hspace{-7.0cm}\delta\phi_{k_m-2}(t=0)=\!\int_0^\infty\!\!d(\Delta
t)\,{\cal P}(\Delta t)\,\delta\phi_{k_m-1}(\Delta
t)\nonumber\\&&\hspace{-7.0cm}\delta\phi_{k_m-1}(t=0)=\!\int_0^\infty\!\!d(\Delta
t)\,{\cal P}(\Delta t)\,\delta\phi_{k_m}(\Delta t)-\!\frac{1}{N}.
\label{e4.15}
\ee      The equation for $\delta\phi_{k_m-1}(\Delta t)$ is different
from the other ones in Eq. (\ref{e4.15}), as it has an extra $-1/N$ on
its r.h.s. This is so, because the one X particle that hopped over to
the $k_m$-th bin at $t=0$, came from the $(k_m-1)$-th bin.

In actuality, Eq. (\ref{e4.15}) should be written in terms of
$\phi_k$'s. If we do so, then on the r.h.s. of the corresponding
equations, we have integrals of the form $\displaystyle{\int_0^\infty
d(\Delta t)\,{\cal P}(\Delta t)\,\phi^{(0)}_k(\Delta t)}$. We have
replaced these integrals by $\phi^{(0)}_{k-1}(t=0)$. This is
consistent with the fact that in an average sense, the underlying
particle density field $\phi^{(0)}_k(t)$ has a uniformly translating
solution. The leftover $\delta\phi_k$ terms then yield
Eq. (\ref{e4.15}).

In terms of this formulation, the leading edge of the front, whose
equation of motion is governed by the linearized equation,
Eq. (\ref{e4.6}), is divided into two parts \cite{notematching}. In
the first part, which lies on the left of (including) the $k_b$-th
bin, the solution is given by the form
$\phi_k(t)=A\sin[z_i(k_m-v_Nt)+\beta]\exp[-z_r(k_m-v_Nt)]$ for
$0<t<\Delta t$. In the second part, constituted by the bins indexed by
$k$, such that $k_b<k\leq k_m$, the shape of front is given by
Eqs. (\ref{e4.8}-\ref{e4.15}). The first part yields the linear
dispersion relation, Eq. (\ref{e3.4}), while the second part yields
more complicated and nonlinear relations between $v_N$, $z_r$ and
$z_i$ involving several other unknown quantities as a self-consistent
set of equations. With the values of $A$, $k_b$ and $k_m$ externally
determined, if one counts the number of equations and the number of
unknowns that are available at this juncture for the selected
asymptotic speed $v_N$, then, from Eqs. (\ref{e3.4}), (\ref{e4.4}),
(\ref{e4.8}) and (\ref{e4.13}-\ref{e4.15}), it is easy to see that
they involve as many unknowns as the number of equations. The value of
$A$ is obtained by matching the mean field solution of the bulk of the
front, where the nonlinearities of Eq. (\ref{e4.5}) play a significant
role, with the solution of the leading edge of the front described by
the linear equations (i.e., Eq. (\ref{e4.6})). On the other hand,
obtaining the value of $k_b$ and $k_m$, for a given set of parameters
$N$ and $\tilde\gamma$, is a more complicated process and now we
address it in the next few paragraphs. We will take up these issues in
further detail in Sec. \ref{sec5d} as well, when we compare our
theoretical results with the results obtained from the computer
simulation.

While it is easy to determine the foremost bin and hence define $k_f$
for any given realization in a computer simulation, the question how
to obtain the values of $k_m$, $\beta$ and $k_b$ for a given set of
values of $N$ and $\tilde\gamma$, still remains to be answered. As a
first step to answer this question, we redefine $A$ and absorb the
quantity $\beta$ in $k$ by a change of variable,
$z_ik+\beta\rightarrow z_ik$, such that in the quasi-comoving frame,
$\phi^{(0)}_k(t)$ reduces to
$A\sin[z_i(k-v_Nt)]\exp[-z_r(k-v_Nt)]$. First, this makes $k$ a
continuous variable as opposed to a discrete integral one. Secondly,
the number of unknown quantities is also reduced from three to two,
namely, to $k_m$ and $k_b$.

If we now look back at Fig. \ref{fig2}, and recapitulate the structure
of the mean-field theory we presented in this section, we realize that
the buildup of particles in the bins at the tip of the front due to
the stalling phenomenon always makes the curve $\phi_k(t=0)$ lie {\it
above\/} $\phi^{(0)}_k(t=0)$, when they are plotted against the
continuous variable $k$. In our mean-field theory,
$\phi_{k_m}(t=0)=1/N$, which clearly means that
$\phi^{(0)}_{k_m}(t=0)<1/N$ and since $\phi^{(0)}_{k}(t=0)$ is a
monotonically decreasing function of $k$, this further implies that
$k_m>k_{m_0}$, where $\phi^{(0)}_{k_{m_0}}(t=0)=1/N$.

In our mean-field theory, what is the numerical value of
$(k_m-k_{m_0})$, the distance between the bin, where the lowest order
approximation $\phi^{(0)}$ reaches the values $1/N$ and the bin, where
the actual average profile $\phi$ reaches this value? For arbitrary
values of $N$ and $\tilde\gamma$, this is not an easy question to
answer.

To check our theory, in this paper we confine ourselves mostly to the
case of $\tilde\gamma=\,$growth rate$\,=1$, as it is the most
illustrative case to demonstrates the multiple facets of fluctuating
front propagation. For a part of the analysis, we also consider the
$\tilde\gamma=0.1$ case. For such values of $\gamma$, i.e., if
$\tilde\gamma$ is too small ($\tilde\gamma\ll1$), or not too large
($\tilde\gamma\sim1$), the only information that we have at our
disposal to obtain the value of the continuous parameter $k_m$, is the
fact that $k_m>k_{m_0}$. For such values of $\tilde\gamma$, therefore,
the only remaining way to generate the ${\cal P}(\Delta t)$ curve is
to use trial values of $k_m$, for $k_m>k_{m_0}$ in an iterative manner
\cite{recursive} [recall that the value of $k_m$ is needed for the
initial condition, Eq. (\ref{e4.14})]. For such values of
$\tilde\gamma$, the use of the trial values of $k_m$ to generate
${\cal P}(\Delta t)$ also requires the value of $k_m-k_b$ as an
external parameter, which can be chosen to be a few, say $\sim4$ (of
course, this number can be increased to obtain higher degree of
accuracy for the $\delta\phi_k(t=0)$ values). We will take up further
details about it in Sec. V. However, before that, we next discuss two
additional fluctuation effects that have important consequences on the
${\cal P}(\Delta t)$ curve. We also mention here that we have explored
the possibility of a relation between $k_f$, obtained from computer
simulation results, and $k_m$, but due to the fact that $k_f$ has
stochastic fluctuations in time, such a relation does not exist.

\subsection{Additional Fluctuation Effects} \label{sec4c}

Having described the mean-field theory, we are now in a position to
assess its accuracy or validity for the probability distribution
${\cal P}(\Delta t)$ that it generates, before we start to look for
numerical confirmation.  At the end of Sec. \ref{sec4a}, we have
mentioned that the fluctuation of the number of X particles in the
foremost bin plays a very significant role in deciding the nature of
${\cal P}(\Delta t)$.  Such fluctuations are not captured in our mean
field theory, which simply assumes that the number of X particles in
the foremost bin at $t=0$ is $1$ and afterwards the number of the X
particles in it increases through the process of a mean growth. In
particular,  at the end of Sec. \ref{sec4a}  we have  described two
kinds of events that, we now argue, affect the nature of ${\cal
P}(\Delta t)$ for large values of $\Delta t$, compared to the time
scale set by $1/v_N$.

\subsubsection{Few Foremost Bins Are Created Too Fast in a
Sequence}\label{toofast}

The first of these events is that if a few of the new foremost bins
are created relatively fast in a row, then soon there would be a case
of a new foremost bin creation that takes an unusually long
time. Naturally, this gives ${\cal P}(\Delta t)$ a higher value than
what our mean-field theory does for large values of $\Delta t$. The
reason for this is quite simple: the mean growth of the number of X
particles in the foremost bin is exponential in time, which would
indicate that if one describes the growth of the number of X particles
in the foremost bin simply by mean growth, then the probability
distribution ${\cal P}(\Delta t)$ decreases very rapidly for large
$\Delta t$, and clearly that fails to describe the slow decay of
${\cal P}(\Delta t)$ for large $\Delta t$ arising out of this event.

Unfortunately, there is no way to estimate the effect of this event
within the scope of {\it any\/} mean-field theory, since by its sheer
nature, it can only be described by the multi-time correlation
functions of the times required for sequential creations of new
foremost bins. For this reason, we call this event ``correlated
diffusion event'' for later reference. But the physical effect of it
can be expressed in a slightly different manner which is more
conducive for understanding the conditions of applicability of our
mean-field theory. In our mean-field theoretical description, before a
new foremost bin is created, the shape of the front is always the {\it
same\/} mean shape, described by Eqs. (\ref{e4.5}). On the other hand,
if a few of the new foremost bins are created relatively fast in a
row, the leading edge of the front gets more and more elongated while
the number of particles inside the bins in the leading edge does not
get a chance to grow accordingly. Thus, this event creates significant
deviation for the actual front shape from the front shape described by
our mean-field theory.  The magnitude of this deviation, measured by
subtracting the mean-field density of the X particles from the actual
density of X particles inside the bins at the leading edge of the
front, is obviously {\it negative\/}. If we combine this argument with
the fact that on an average, the probability of a new foremost bin
creation increases with the increasing number of X particles in the
foremost bin, then it is easy to realize that after a sequence of such
fast creations of new foremost bins, the front needs to replenish the
number of X particles in the leading edge before another new foremost
bin is created. It is this replenishing process which is responsible
for the next new foremost bin creation at a relatively long time.

It is now intuitively clear that in terms of the front shape, the
larger the deviation such an event causes, the more ${\cal P}(\Delta
t)$ will be affected for large values of $\Delta t$. Based on this, we
now argue that for a fixed value of $N$, such an event does not affect
the large $\Delta t$ behaviour of ${\cal P}(\Delta t)$ curve for large
values of $\tilde\gamma$ as much as it does for small values of
$\tilde\gamma$. To reach this conclusion, one simply needs to observe
the following: the mean shape and the corresponding density of the X
particles in the bins at the leading edge of the front is
characterized by $z_r$ and $z_i$, and for small $\tilde\gamma$, the
values of $z_r$ and $z_i$ is large and vice versa (as $z_i\sim
z_r\sim\tilde\gamma^{-1/2}$, see Eq. (\ref{e3.11})). For large
$\tilde\gamma$, therefore, for the mean shape of the front, the
leading edge is already quite elongated and the density of the X
particles at the tip of the front is quite small, compared to their
small $\tilde\gamma$ values. As a result, for small $\tilde\gamma$,
the magnitude of the deviation from the mean front shape, caused by
such an event is much larger, and since the growth rate is always
unity for all $\tilde\gamma$, it takes a much longer time to replenish
the density of X particles for small $\tilde\gamma$ than for large
$\tilde\gamma$ values.

\subsubsection{Particles in the Foremost Bin Jumps Back}\label{backjump}
\label{sec4c2}

The other kind of fluctuation effect has to do with the fact that
albeit according to our definition, the foremost bin remains the
foremost one until time $t=\Delta t$, it may so happen that at some
nonzero value of $t$, all the X particles diffuse back to the left
leaving the foremost bin empty for some time and then another X
particle diffuses into the foremost bin from the left, making it
non-empty again at a finite value of $t$, say at $t=t_0$. Clearly,
this event is much more unlikely to take place once the number of X
particles in the foremost bin has grown, since in that case, all the X
particles in the foremost bin have to diffuse back to the
left. Essentially, this event is therefore restricted to the following
sequence: {\em (a) } starting $t=0$, the foremost bin remains occupied
by a single X particle for sometime, {\em (b)} this X particle then
diffuses back to the left leaving the foremost bin empty, until {\em
(c)} another X particle diffuses into the foremost bin, making it
non-empty again at $t=t_0$. Of course, the value of $t_0$ is not fixed
and it is chosen probabilistically. For later reference, we call this
``the vacant foremost bin event'', and this event is much more likely
to take place for large diffusion coefficient $\tilde\gamma$.

Based on the picture described in {\em (a)-(c)}  above in the previous
paragraph, we can now make a quantitative estimate of this particular
event and accordingly correct the expression of ${\cal P}(\Delta
t)$. One simply has to realize that if  this event takes place, then
the time at which the theory for mean growth of X particles in the
foremost bin (with exactly one X particle to start with) can be
applied in this quasi-comoving frame, shifts from $t=0$ to
$t=t_0$. However,  we also need to obtain an estimate for the value of
$t_0$. This can be obtained using the following argument: if in step
{\em (b)}, the only X particle in the foremost bin had diffused to the
right, instead of diffusing to the left, it would have been a case of
a new foremost bin creation, the time scale for which is set by
$1/v_N$. Since the probability of this single X particle in the
foremost bin to jump to the right is the same as the probability of it
to jump to the left, we can also say that the time it takes for the X
particle in the foremost bin to diffuse back to the left takes
approximately a time $1/v_N$ starting $t=0$. Similarly, step {\em (c)}
is exactly the same step as a ``new foremost bin creation''. Hence,
after step {\em (b)}  is over, it takes a further $1/v_N$ time
\cite{timescale} for another X particle to diffuse from the left into
the foremost bin. Together, these two events make $t_0\approx2/v_N$
\cite{2/v_N}, and this argument illustrates that this event affects
the behaviour of ${\cal P}(\Delta t)$ only for $\Delta
t\gtrsim2/v_N$. Having neglected the effect of the correlated
diffusion events on the ${\cal P}(\Delta t)$ curve for $t>2/v_N$ (for
which we have no theoretical estimate anyway), if we now claim that
for {\it all\/} $\Delta t$ values greater than $2/v_N$, the population
of the X particles in the foremost bin is described by
Eqs. (\ref{e4.5}), but with the condition that $\langle
N_{k_m}(t=t_0)\rangle=N_{k_m}(t=t_0)=1/N$, as opposed to having
$\langle N_{k_m}(t=0)\rangle=N_{k_m}(t=0)=1/N$, then we can still
incorporate the effect of this event (that arise out of fluctuations)
within the scope of the mean-field theory that we described in this
section. If this procedure is correct, then while comparing the
theoretical ${\cal P}(\Delta t)$ curve with the ${\cal P}(\Delta t)$
curve obtained from the simulations, one would notice that for large
values of $\Delta t$, this procedure underestimates the magnitude of
${\cal P}(\Delta t)$. Hitherto, this underestimation then would be an
indication of the effect of the correlated diffusion events on ${\cal
P}(\Delta t)$. We will return to these points once again in the next
section, where we seek numerical confirmation of our theory
\cite{highd} \label{2/v_Npage}.

\subsection{Summary of the Status of the Present Approach and Additional
Observations \label{4D}}

The discussion above completes the theoretical formulation for the
asymptotic speed selection of the front. Before we discuss how given
values of $N$ and $\tilde\gamma$ would generate the corresponding
values of $v_N$ from our mean-field theory described in this section,
we summerize our claims here and make a number of additional
observations.

{\em 1)} Based upon the microscopic description of the front movement,
we have formulated a mean-field theory that describes, on a lattice,
the front propagation as a sequence of ``halt-and-go'' process. In
this way of looking at the front propagation, essentially the number of
X particles at the tip of the front determines the asymptotic speed of
the front. Since the number of X particles at the tip of the front are
rather few, the fluctuations in the number of X particles at the tip
of the front affects the asymptotic speed of the front in a strong
manner. Part of the fluctuation effects can be estimated within the
scope of this mean-field theory itself. The other part, for which the
fluctuations can only be studied by means of a multi-time correlation
functions, is expected to affect the accuracy of our theory much more
for small $\tilde\gamma$ than for large $\tilde\gamma$
values. Therefore, overall, in terms of numerical confirmation, one
can expect to find a greater accuracy for large values of
$\tilde\gamma$.

Moreover, for large $\tilde\gamma$, the discreteness of the lattice
effects are suppressed, and therefore, for a given value of $N$, one
would expect that the relative correction for the asymptotic front
speed, $(v^*-v_N)/v^*$, must become small.

{\em 2)} There are two important aspects that one must  take notice
of. First, in a mean field description that  incorporates the effect
of the stalling phenomenon, we have demonstrated from the microscopic
dynamics that there {\it exists a cutoff of particle density\,}, which
is expressed by the fact that in this mean field description,
$\phi_k(t)=0$ for $k>k_m$ against a finite value at $k=k_m$.
Secondly, we have also demonstrated that the quantity $a$ in Sec. III
is indeed an effective quantity only,  as the solution of the
linearized equation of the front, given by Eq. (\ref{e3.3}), is not
valid near the foremost bin, and the fact that the asymptotic speed
selection mechanism arises  from a proper probabilistic description of
the tip of the front.

{\em 3)} For very small values of $\tilde\gamma$, we have previously
noticed that the correlated diffusion event plays a very dominant role
that no mean-field theory can ever generate, so we should leave the
$\tilde\gamma\ll1$ case outside the purview of our mean-field theory
(we will demonstrate this in the next section).

{\em 4)} To judge the appropriateness of our mean-field theory, as far
as the generation of the numerical value of the asymptotic front speed
for given values of $N$ and $\tilde\gamma$ is concerned, we make the
following observations: {\em (i)} the case of $\tilde\gamma\ll1$
cannot be studied in terms of a mean-field theory, {\em (ii)} the case
of $\tilde\gamma\sim1$ needs a trial value of $k_m>k_{m_0}$ and the
use of a recursive feedback mechanism to generate the ${\cal P}(\Delta
t)$ curve, and {\em (iii)} we still need the values of $A$, $z_r$ and
$z_i$, which can be obtained only from the simulation data for given
values of $N$ and $\tilde\gamma$. In view of these points, it is clear
that this theory is unable to make a definitive predictions for $v_N$,
without any assistance from the computer simulations
whatsoever. Moreover, the Eqs. (\ref{e4.4}-\ref{e4.15}), which one
needs to solve to generate the ${\cal P}(\Delta t)$ curve, are highly
nonlinear equations, hence, this theory can only hope to {\it show
consistency\/} with the results of the computer simulations, as
opposed to produce a numerical value of $v_N$ which is then
subsequently confirmed by the computer simulations.

{\em 5)} Finally, we note that unlike Eq. (\ref{e3.12}), this theory
does not make the effect of the value of $N$ on the asymptotic front
speed explicit. However, it is natural to expect that the effect of
stalling of the front and the associated particle density buildup at
the tip of the front on the front shape and speed would become less
and less for increasing $N$. This would reflect in the comparison of
the $\delta\phi_k(t=0)$ values against the corresponding
$\phi_k^{(0)}(t=0)$ values in the bins at the very tip of the
front. We would return to this point in Sec. \ref{VD1}.

\section{Test of the Theory against Computer Simulations}

We now check our theory, as it has been presented in Secs. III and IV,
against the results of the computer simulations. There have been quite
a few aspects of the theory that we have presented in Secs. III and
IV; and for a given set of values of $N$ and $\tilde\gamma$, testing
all these aspects of our theory is not a short and easy process. To
explain how we do the simulations, obtain $v_N$ and $A$, and check the
front shape, we choose one particular set of $N$ and $\tilde\gamma$
values, namely $N=10^4$ and $\tilde\gamma=1$. We then use these
methods to obtain the simulation data for three other values of $N$,
namely $N=10^2$, $N=10^3$ and $N=10^5$, keeping the value of
$\tilde\gamma$ fixed at $1$. Based on this scheme, this section is
divided into five subsections. In Sec. \ref{sec5a}, we present the
simulation algorithm and obtain $v_N$ for $\tilde\gamma=1$ and
$N=10^4$. In Sec. \ref{Asection}, we summarize the method to calculate
$A$, and subsequently obtain its value for $\tilde\gamma=1$ and
$N=10^4$ using the results of Sec. \ref{sec5a}. In the
Sec. \ref{sec5c}, we contrast the simulation results of
Secs. \ref{sec5a} and \ref{Asection} with the theory of Sec. III. In
Sec. \ref{sec5d}, we test our theoretical predictions for ${\cal
P}(\Delta t)$ against the computer simulation results for $N=10^4$
$10^2$, $10^3$ and $10^5$, and $\tilde\gamma=1$ (in that
order). Moreover, in Sec. IV, we have conjectured that the mean-field
theory mimicking the stalling phenomenon would be less successful for
small values of $\tilde\gamma$. We verify this conjecture in
Sec. \ref{VD1} by means of a relative comparison of the theoretical
and simulation ${\cal P}(\Delta t)$ curves for $\tilde\gamma=0.1$ and
$N=10^4$. We also remind the reader that in
Secs. \ref{sec5a}-\ref{sec5c}, $k$ and $t$ respectively denote the
laboratory bin co-ordinate and actual physical time (and therefore
they do {\it not\/} relate to the quasi-comoving co-ordinates or the
resetting of clocks that requires $0<t<\Delta t$).

\subsection{Computer Simulation Algorithm \label{sec5a}}

Our algorithm for carrying out the computer simulations is the same as
it has been described in Ref. \cite{breuer}. The starting density
profile  of the X particles is a step function, given by
$\phi_k(t=0)=[\,1-\Theta(k-k_0)\,]$, for some $k_0$. The simulation
algorithm consists of a repetitive iterations of two basic steps:\\
{\em (i)}  Let us assume that at any time $t$, the configuration of
the system is given by $(N_1,N_2,\ldots,N_{k'})$, for some $k'$. The
total rate of possible transitions, $W_k$, for the $N_k$ number of X
particles in the $k$-th bin are the sum of $2N_k$ diffusions, creation
of $N_k$ new X particles and annihilation of $N_k(N_k-1)/N$ number of
X particles, i.e.,

\be  W_k\,=\,2\tilde\gamma\,N_k\,+\,N_k\,+\,\frac{N_k(N_k-1)}{N}\,.
\label{e5.1}
\ee  The total rate of transition, $W_{tot}$, for all the bins is
therefore   \be  W_{tot}\,=\,\sum_{k=1}^{k'}\,W_k\,.
\label{e5.2}
\ee  Starting at time $t$, the probability of no transition happening
for an interval $\tau$ is given by   \be
\wp(\tau)\,=\,\exp(-\,W_{tot}\,\tau)\,.
\label{e5.3}
\ee   Before any transition takes place, a random number $r_0$ is
chosen within the interval $[0,1)$. The time $\tau$ that one needs to
wait before any transition happens is then determined as   \be
\tau\,=\,-\,\frac{1}{W_{tot}}\,\ln r_0\,.
\label{e5.4}
\ee
\noindent {\em (ii)} With the time $\tau$ for a transition at our
disposal,  the bin where the transition takes place and the specific
transition in that bin must also be determined. To do so, we choose
another set of two random numbers, $r_1$ and $r_2$, in $[0,1)$. From
the numerical value of $r_1$ and the fact that the probability of a
transition taking place in the $k$-th bin is given by $W_k/W_{tot}$,
we determine the index of the bin where the transition takes
place. Similarly, the particular transition in the $k$-th bin is
determined from the numerical value of $r_2$ and considering the
probabilities of different kinds of transitions in the $k$-th bin: \be
\mbox{probability of a diffusion to the
right}\,=\,\frac{\tilde\gamma\,N_k}{W_k}\,,\nonumber\\&&\hspace{-7.2cm}\mbox{probability
of a diffusion to the
left}\,=\,\frac{\tilde\gamma\,N_k}{W_k}\,,\nonumber\\&&\hspace{-7.2cm}\mbox{probability
of breeding a new X
particle}\,=\,\frac{N_k}{W_k}\,,\,\,\mbox{and}\nonumber\\&&\hspace{-7.2cm}\mbox{probability
of annihilating an X
particle}=\frac{N_k(N_k-\!1)}{NW_k}.\nonumber\\&&\hspace{-7.2cm}
\label{e5.5}
\ee
\begin{figure}
\begin{center}
\includegraphics[width=2.9in,angle=270]{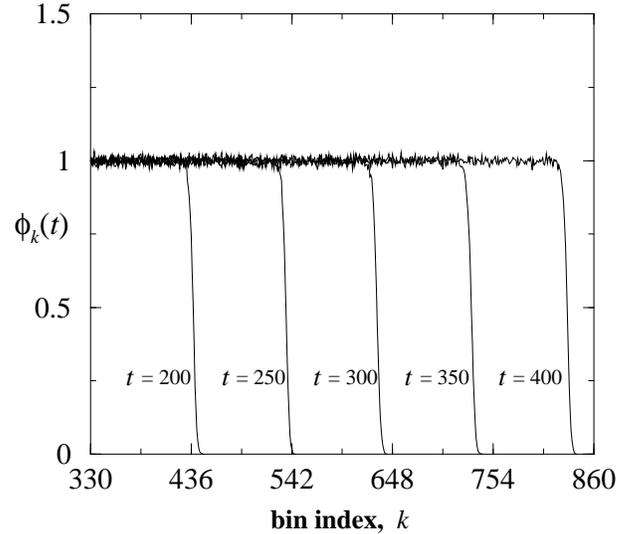}
\caption{To illustrate that the front reaches its steady state shape
before $t=200$, the plot of $\phi_k(t)$ vs. $t$ for five different
values of $t$ spaced at regular intervals, $t=200,\,t=250,\, t=300,\,
t=350$ and  $t=400$, are shown above. \label{fig3}}
\end{center}
\end{figure}
\noindent Once the transition is determined, the configuration of the
system is subsequently updated. However, any X particle diffusing from
the first bin (i.e., $k=1$) towards the left is immediately replaced.

In this subsection, we focus on one particular set of values of
$\tilde\gamma$ and $N$, namely, $\tilde\gamma=1$ and $N=10^4$. The
value of $k_0$ for the initial density profile of the X particles is
chosen to be 50. Starting at $t=0$, we let this initial profile evolve
in time. To obtain the random numbers, we use the random number
generator ${\mathsf drand48}$ provided in the standard C library
functions with the initial seed \cite{seed} $s=123456$. It turns out
that to a very good approximation, the front shape reaches a steady
state somewhere before $t=200$. The front shapes from $t=200$ to
$t=400$ is shown  in Fig. \ref{fig3} as an illustration. For
measurement of the asymptotic quantities, therefore, we take $t=200$
as our starting point.

\begin{figure}[h]
\begin{center}
\hspace{-8mm}\includegraphics[width=2.4in,angle=270]{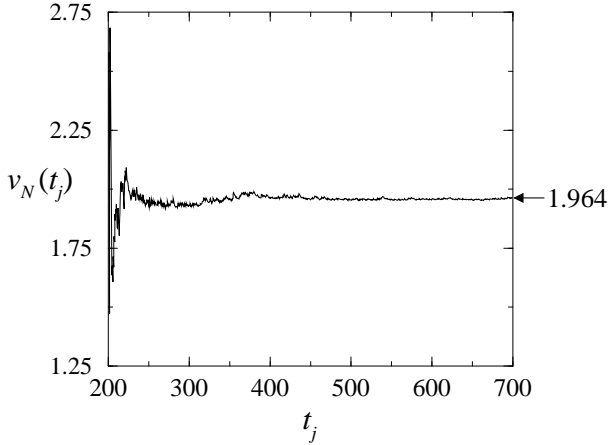}
\caption{Values of $v_N(t_j)$ for $200<t_j<700$ and
$j=1,2,\ldots,980$. As expected, the fluctuations in $v_N(t_j)$ die
out for large values of $j$.\label{fig4}}
\end{center}
\end{figure}

To calculate the asymptotic speed of the front, we measure the $\Delta
t$ values for creating new foremost bins after $t=200$ till
$t=700$. We find that altogether there are $980$ different $\Delta t$
values in this time interval. Assuming that the $j$-th value of
$\Delta t$ takes place at time $t_j$ ($j=1,2,\ldots,980$), we define
the $j$-th cumulative average of the $\Delta t$ values as \be
\langle\Delta t\rangle_j\,=\,\frac{1}{j}\,\sum_{j'=1}^{j}\Delta
t_{j'}\,,
\label{e5.6}
\ee which subsequently allows us to define the speed at time $t_j$ as
\be v_N(t_j)\,=\,\frac{1}{\langle\Delta t\rangle_j}\,.
\label{e5.7}
\ee  Naturally, for small values of $j$, the values of $v_N(t_j)$
fluctuate, but as $j$ becomes large, the fluctuations die out and
$v_N(t_j)$ approaches $v_N$. The plot of $v_N(t_j)$ vs. $t_j$ is
shown in Fig. \ref{fig4} for $j=1,2,\ldots,980$, $t_1=200.562$ and
$t_{980}=699.271$. We notice from the plot that the fluctuations in
$v_N(t_j)$ are really small for $t_j>500$. The $v_N(t_j)$ values for
$t_j>500$, therefore, allows us to set the error bar on the
measurement of $v_N$ in Eq. (\ref{e5.8}), and we obtain  \be
v_N\,=\,1.964\,\pm\,0.006\,.
\label{e5.8}
\ee

\subsection{The coefficient  $A$  as a reflection of the nonlinear
front behavior\label{Asection}}

The quantity $A$ has been introduced to solve for the linear
difference-differential equation, Eq. (\ref{e3.0}). Its numerical
value, however, cannot be determined from the linear equation, since
{\it any\/} value of $A$ satisfies it. To determine the value of $A$,
therefore, one needs to solve the full {\em nonlinear}
difference-differential equation, Eq. (\ref{e2.6}), expressed in terms
of the comoving coordinate, $\xi$. This is done, together with the
associated values of the real and imaginary part of $z$,  in Appendix
\ref{Aappendix}; for $\tilde{\gamma}=1$ and $N=10^4$, we find
\begin{equation}
A=0.961 \pm 0.012 \, .\label{e5.14b}
\end{equation}

\subsection{Numerical Test  of the Predictions for Consistency of
Front Shape and Speed \label{sec5c}}

Equipped with the value of $A$ and the simulation data, we are now in
a position to contrast the result of Sec.~III with the simulation
results. The purpose of this subsection is twofold: first, we
demonstrate that the theoretical shape of the front generated in
Sec. \ref{Asection} for $\tilde\gamma=1$ and $N=10^4$ agrees very well
with the shape of the front obtained from the simulation data by
taking snapshots at arbitrary times. Secondly, we demonstrate that
there are {\it significant\/}  differences in the two values of
$v^*-v_N$, one obtained from Eq. (\ref{e3.14}), and the other from
Eq. (\ref{e5.8}).
\begin{figure}[h]
\begin{center}
\includegraphics[width=2.9in,angle=270]{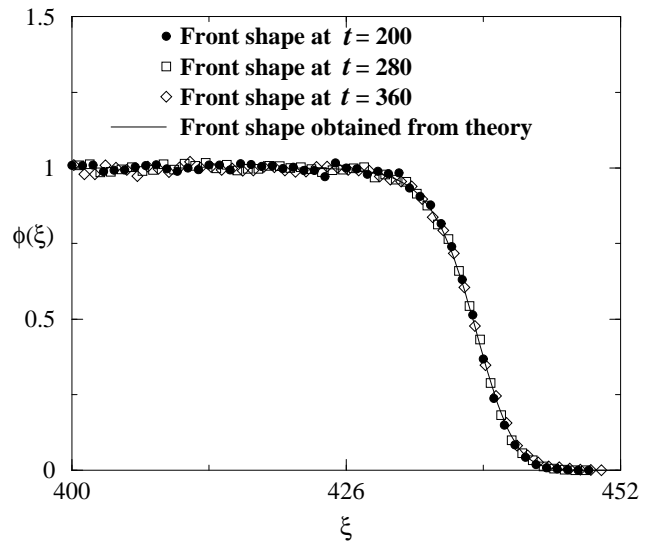}
\caption{Theoretical shape of the front represented by the solid line
and the shape of the front obtained from computer simulations at three
discrete times, $t=200$, $t=280$ and $t=360$, represented by three
different symbols.
\label{fig5}}
\end{center}
\end{figure}
To compare the shapes of the front obtained from the theory and the
simulations, we choose to take snapshots at three discrete times,
$t=200$, $t=280$ and $t=360$. Below we present the plot that contains
the front shapes for these times obtained from the simulations and the
theory. The comoving coordinate $\xi$ for the bins is chosen in a way
such that it coincides with the laboratory frame position of the front
at $t=200$. As we can see from the graph below, the collapse of the
data for $t=200$, $t=280$ and $t=360$ is very good and the solid line
representing the theoretical prediction is almost indistinguishable
from the simulation data. The theoretical curve is first generated
using the method described in Sec. \ref{Asection} with $v_N(t_{980})$
as the asymptotic front speed, the corresponding values of $z_l$,
$z_r$, $z_i$ and $A$ of Eqs. (\ref{e5.14}) and (\ref{e5.14b}), and
then having it shifted to coincide with the laboratory frame position
of the front at $t=200$.

We now return to the result of Sec. III, and denote $v^*-v_N$ obtained
from Eq. (\ref{e3.14}) by $\Delta v_{asymp} $. For $\tilde\gamma=1$,
Eq. (\ref{e3.1}) yields $z_0=0.9071032..$ and $v^*=2.07344$. Using
these values for $N=10^4$, we obtain  \be \Delta
v_{asymp}\,=\,0.152024\ldots\,,
\label{e5.16}
\ee On the other hand, using $v_N(t_{980})$ for $v_N$, the value
$\Delta v_{sim} = v^*-v_N$ comes out to be \be \Delta v_{sim}
\,\equiv\,v^*\,-\,v_N(t_{980})\,\approx\,0.11\,,
\label{e5.17}
\ee which implies that the asymptotic estimate $\Delta v_{asymp}$ is
about $38\%$ larger than $\Delta v_{sym}$  from the computer
simulations.

These results clearly indicate that for large but not extremely large
values of $N$ ($N=10^4$ here), there is much more to the story than
$v^*-v_N$ being simply $\propto\ln^{-2}N$. The theory presented in
Sec. III does capture the essentials, and it would have been good
enough to generate appropriate numbers for $v^*-v_N$, if one could
obtain the value of $a$ externally. However, in view of the fact that
the uniformly translating solution of Eq. (\ref{e2.6}) cannot be
extended all the way up to the foremost bin, the quantity $a$ is a
fictitious and simply an effective quantity (already mentioned in
Sec. III). Therefore, it is not possible to obtain the numerical value
of $a$ from computer simulation results or from any theoretical
estimate. Besides, the theory of Sec. III completely overlooks the
microscopic intricacies at the tip of the front, and hence, it should
be regarded as an effective theory.

\subsection{Numerical Test of the Theory for ${\cal P}(\Delta t)$
\label{sec5d}}

In this subsection, we seek the numerical test of our theory presented
in Sec. IV. We carry out this task in two steps. In the first step, we
check most of the aspects of the theory for $N=10^4$ and
$\tilde\gamma=1$, where we describe the method for obtaining ${\cal
P}(\Delta t)$. Subsequently, in the second step, we check the
predictions of our theory for $N=10^2$, $10^3$ and $10^5$, keeping the
value of $\tilde\gamma$ fixed at $1$. {\it Notice that comparing
probability distributions allows us to verify more detailed
representations of the actual forward movements of the foremost bin
against comparing only the asymptotic front speed $v_N$, which is the
inverse of the first moment of ${\cal P}(\Delta t)$ [see
Eq. (\ref{e4.4})]\/}. We should note that in view of the strong
nonlinearity of the self-consistent theory of Sec. IV, we will have to
use the values of $A$, $z_r$, $v_N$ and $z_i$ obtained from computer
simulations to generate the ${\cal P}(\Delta t)$ curve, and then
obtain the theoretical value of $v_N$. This process therefore becomes
a self-consistency check of our theory of Sec. IV, as opposed to a
verification of its predictions. Moreover, we do not compare the
${\cal P}(\Delta t)$ curves directly. This is for a very simple
reason: namely that the expression for ${\cal P}(\Delta t)$ in
Eq. (\ref{e4.8}) involves $\langle N_{k_m}(\Delta t)\rangle$ as a
coefficient. In an actual computer simulation, this quantity
fluctuates wildly, and hence, generating a histogram to obtain the
probability distribution ${\cal P}(\Delta t)$ from computer
simulations proves to be difficult. Instead, we compare the
``cumulative probability distribution'' curves $P(\Delta t)$, which is
defined as the probability of a new foremost bin creation happening at
time $t\geq\Delta t$. From a theoretical point of view, the expression
of $P(\Delta t)$ can be found easily from Eq. (\ref{e4.8}) as \be
P(\Delta t)\,=\,\int_{\Delta t}^{\infty}dt'\,{\cal
P}(t')\nonumber\\&&\hspace{-2.4cm}=\,\exp\!\left[-\,\tilde\gamma\,\int_0^{\Delta
t}\!dt\,\langle N_{k_m}(t)\rangle\right]\,.
\label{e5.18}
\ee It turns out the $P(\Delta t)$ histogram generated from the
computer simulation results is not noticeably affected by
fluctuations, which makes its comparison with the $P(\Delta t)$ curve
generated from our theory much simpler.

\subsubsection{The Case of $\tilde\gamma=1$ and $N=10^4$}

\begin{figure}[h]
\begin{center}
\includegraphics[width=2.8in,angle=270]{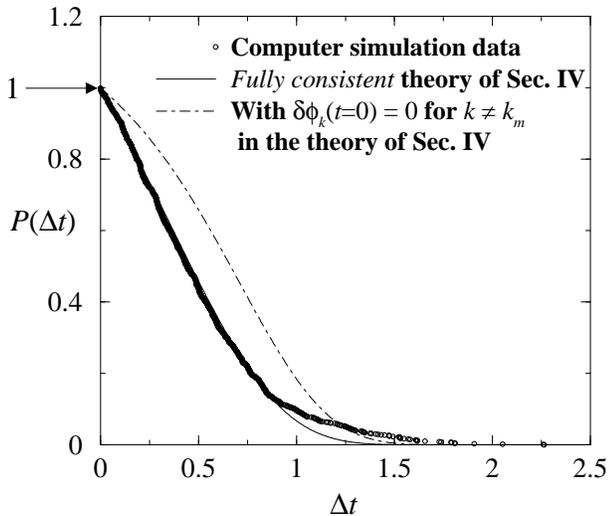}
\caption{The cumulative probability distribution $P (\Delta t)$ as a
function of $\Delta t$ for $N=10^4$ and $\tilde\gamma=1$. \label{fig6}}
\end{center}
\end{figure}

The $P (\Delta t)$ curve from the computer simulations are generated
in the following way: by definition $P(0)=1$. For $N=10^4$ and
$\tilde\gamma=1$, there are $980$ values of $\Delta t$. First, these
are arranged in an increasing order of magnitude, $\Delta t_1, \Delta
t_2,\ldots, \Delta t_{980}$, and then in the corresponding values of
$P(\Delta t)$ are obtained as \be P(\Delta t_{j+1})\,=\,P(\Delta
t_j)\,-\,\frac{1}{980}
\label{e5.19}
\ee for $j=2, 3, \ldots, 980$ with the initial condition that
$P(\Delta t_1)=1-1/980$. The corresponding $P(\Delta t)$ vs. $\Delta
t$ plot is shown in Fig. \ref{fig6} by open circles.

To generate the corresponding theoretical cumulative probability
distribution, we proceed in the following way. {\it In a coordinate
system,  where the function $\phi^{(0)}_k(t=0)=0$ at $k=\pi/z_i$\/}
[which  allows us to use the values of $A$, $z_r$, $v_N$ and $z_i$ of
Eq. (\ref{e5.14})], we work out the whole machinery described by Eqs.
(\ref{e4.5}-\ref{e4.15}) {\it neglecting the fluctuation effects
described in Sec. \ref{sec4c}}. This process requires the value  of
$k_m-k_b$, i.e., the number of bins at the tip of the front where  the
buildup of particle density is significant, as well as the value of
$k_m-k_{m_0}$ as external inputs, and we choose $k_m-k_b=4$ for this
purpose \cite{linear} (we refer back to Table~I for the definitions of
$k_{m_0}$ etc.). The calculation of the value of $k_m-k_{m_0}$ and the
generation of the ${\cal P}(\Delta t)$ curve are carried out
self-consistently and hence by iteration, using the recursive feedback
method \cite{recursive}. However, to generate the ${\cal P}(\Delta t)$
curve for any guess value of $k_m-k_{m_0}$, one still needs to have
the values of $\delta\Phi(t=0)$ for the $k_m-k_b$ bins at the tip as a
starting point [see Eq. (\ref{e4.13})]. At the same time, we notice
that $\delta\Phi(t=0)$ can only be determined once the probability
distribution ${\cal P}(\Delta t)$ is obtained. We choose to address
this problem the following way: for any guess value of $k_m-k_{m_0}$,
we start with Eq. (\ref{e4.14}) and $\delta\phi_k(t=0)=0$ for the rest
of the $k_m-k_b$ bins. Keeping $k_m-k_{m_0}$ fixed, we then generate
the corresponding ${\cal P}(\Delta t)$ curve and obtain the
$\delta\phi_k(t=0)$ values for the rest of the $k_m-k_b$ bins using
Eq. (\ref{e4.15}). We keep repeating this process until we converge in
terms of the $\delta\phi_k(t=0)$ values, i.e., when the recursive
correction to the values of $\delta\phi_k(t=0)$ becomes less than
$10\%$ of the $\delta\phi_k(t=0)$ values at the previous step in the
recursion. Once this point is reached for a value of $k_m-k_{m_0}$, we
then compare the theoretical cumulative probability distribution
$P(\Delta t)$ with Fig. \ref{fig6} above to decide upon the next guess
value of $k_m-k_{m_0}$ in the recursive feedback method.

For $N=10^4$ and $\tilde\gamma=1$, the value of $k_m-k_{m_0}$ turns
out to be $k_m-k_{m_0}=1.1431$. We present the corresponding
theoretical cumulative probability curves in Fig. \ref{fig6}. The
solid line in Fig. \ref{fig6} represents the {\it fully consistent\/}
solution of Eqs. (\ref{e4.5}-\ref{e4.15}), while the dashed line
represents the theoretical cumulative probability curve obtained by
solving Eqs. (\ref{e4.5}-\ref{e4.15}) with $\delta\phi_k(t=0)=0$ for
$k\neq k_m$. The fact that the fully consistent solution curve matches
the computer simulation one much better than the naive approximation
where all the $\delta\phi_k$ corrections behind the foremost bin are
ignored is a strong indication of how significantly the buildup of
particle densities in the bins {\it behind the foremost one\/} affects
the property of $P(\Delta t)$.
\begin{figure}[h]
\begin{center}
\includegraphics[width=2.9in,angle=270]{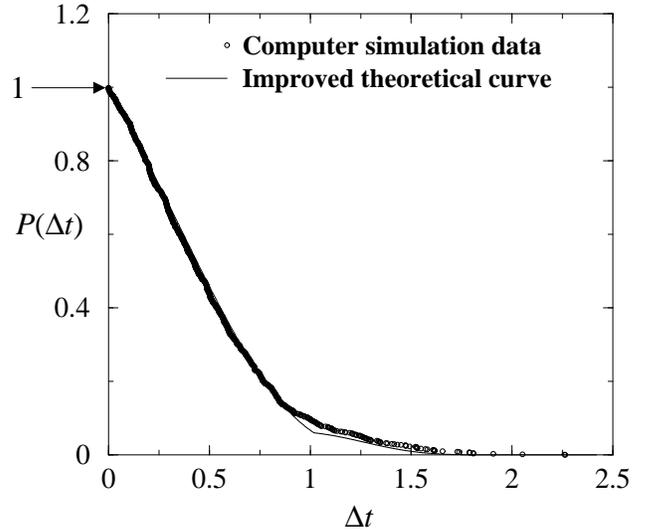}
\caption{The theoretical curve, which includes the effect of the
vacant foremost bin events, and the simulation data for the cumulative
probability distribution $P (\Delta t)$ are presented above for
$N=10^4$ and $\tilde\gamma=1$. \label{fig7}}
\end{center}
\end{figure}
An examination of Fig. \ref{fig6} immediately reveals that the
agreement between the $P(\Delta t)$ curve generated by the fully
consistent theory of Sec. IV and the computer simulation is extremely
accurate up to about $\Delta t=0.9$. However, the theory is unable to
capture the ``tail'' of the $\mbox{\bf P}(\Delta t)$ curve for large
$\Delta t$. Analysis of our data shows that this is due to the
fluctuation effects discussed in Sec. \ref{sec4c}. As mentioned there,
correlated diffusion events are not captured in this theory. However,
we can follow the argument of Sec. \ref{sec4c2} to take into account
the effect of the vacant foremost bin events on the $P(\Delta t)$
curve for large $\Delta t$ values: we assume that all cases of $\Delta
t>t_0\approx2/v_N$ are due to the vacant foremost bin events. This
means that for $\Delta t<t_0$, the $P(\Delta t)$ curve is given by the
solid line in Fig. \ref{fig6}, but from $t_0$ onwards, the $P(\Delta
t)$ curve must be generated from the mean-field dynamics of the tip
described in Sec. IV, with the same value of $k_m$, but with the
initial condition $\langle N_{k_m}(t=t_0)\rangle=N_{k_m}(t=t_0)=1/N$,
as opposed to $\langle N_{k_m}(t=0)\rangle=N_{k_m}(t=0)=1/N$. With the
value of $k_m$ already determined in this subsection, the
corresponding equations, Eqs. (\ref{e4.5}-\ref{e4.15}), are easy to
solve self-consistently as before. From this analysis, we obtain the
behaviour of $P(\Delta t)$ for $\Delta t>t_0$, having noticed that
$P(\Delta t\rightarrow t_0+)$ must be the same as the value obtained
from the solid line in Fig. \ref{fig6} at of $\Delta t=t_0$, i.e.,
$0.06034$. We present the final theoretical $P(\Delta t)$ curve
together with the simulation data in Fig. \ref{fig7}. Notice that this
process introduces a finite discontinuity in the density of the X
particles inside the foremost bin at $\Delta t=t_0$, since for $\Delta
t<t_0$, the density of the X particles in the foremost bin is obtained
from a fully consistent theory of Sec. IV, while {\it at\/} $\Delta
t=t_0$, it is set equal to $1/N$ manually. As a consequence, the
theoretical $P(\Delta t)$ curve in Fig. \ref{fig7} has a slope
discontinuity at $\Delta t=t_0$.

In Fig. \ref{fig7}, the improved theoretical curve follows the curve
reasonably well, but it still lies {\it below\/} the simulation data
points for $\Delta t\gtrsim2/v_N$, as it should be. This discrepancy
gives us an idea about the effect of the correlated diffusion events
on the $P(\Delta t)$ curve that we could not estimate. Using
Eq. (\ref{e5.7}) to calculate the front speed from the theoretical
curve in Fig. \ref{fig7}, we obtain \be
v_N(\mbox{theoretical})\,=\,1.98882\,.
\label{e5.20}
\ee  This is about 0.024 higher than the asymptotic front speed
measured by the computer simulation [see Eq. (\ref{e5.8})], in
agreement with the fact that the theoretical curve for $P(\Delta t)$
slightly underestimates the simulation one for $\Delta t\gtrsim2/v_N$.

\subsubsection{The Cases of $N=10^2$, $10^3$ and $10^5$, with
$\tilde\gamma=1$ \label{VD1}}

\begin{figure}[h]
\begin{center}
\includegraphics[width=2.7in,angle=270]{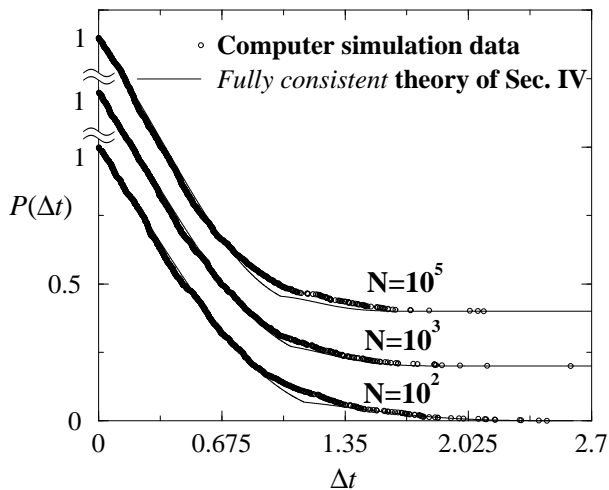}
\caption{The combined theoretical curve and the simulation data for
the cumulative probability distributions $P (\Delta t)$ vs. $\Delta t$
for $N=10^2$, $10^3$ and $10^5$, and $\tilde\gamma=1$. The curves for
the latter two are shifted upwards for clarity. \label{fig7a}}
\end{center}
\end{figure}

We now further test our theory for $N=10^2$, $10^3$ and $10^5$,
keeping the value of $\tilde\gamma$ fixed at 1. The values of
$v_N\mbox{(simulation)}$, $z_r$, $z_i$ and $A$ in Table II below. The
corresponding $P(\Delta t)$ vs. $\Delta t$ graphs, which are the
analogs of the graph in Fig. \ref{fig7}, have been plotted together in
Fig. \ref{fig7a}. Table II presents the theoretical values of $v_N$
that are calculated using these $P(\Delta t)$ vs. $\Delta t$ graphs,
and predicted $v_N$ from Eq. (\ref{e3.14}).  \bleq
\begin{center}
\begin{tabular}{||@{}c|c@{}|c@{}|c@{}|c@{}|c@{}|c@{}||}
\hline\hline
\rule[-2mm]{0pt}{4ex}\,\,\,\,$N$\,\,\,\,&\,\,\,\,$v_N\mbox{(simulation)}$
\,\,\,\,&$z_r$&$z_i$&$A$&\,\,\,\,$v_N\mbox{(theoretical)}$\,\,\,\,&\,\,\,\,$v_N\mbox{[Eq.
(\ref{e3.14})]}$\,\,\,\,\\  \hline
\rule[-2mm]{0pt}{4ex}$10^2$&\,\,\,\,$1.778$\,\,\,\,&
\,\,\,\,$0.8217$\,\,\,\,&\,\,\,\,$0.436$\,\,\,\,&
\,\,\,\,$0.8836$\,\,\,\,&\,\,\,\,$1.808$\,\,\,\,&$1.465$\\ \hline
\rule[-2mm]{0pt}{4ex}$10^3$&\,\,\,\,$1.901$\,\,\,\,&
\,\,\,\,$0.8586$\,\,\,\,&\,\,\,\,$0.3313$\,\,\,\,&
\,\,\,\,$0.9042$\,\,\,\,&\,\,\,\,$1.899$\,\,\,\,&$1.803$\\ \hline
\rule[-2mm]{0pt}{4ex}$10^5$&\,\,\,\,$2.001$\,\,\,\,&
\,\,\,\,$0.8885$\,\,\,\,&\,\,\,\,$0.2654$\,\,\,\,&
\,\,\,\,$1.0714$\,\,\,\,&\,\,\,\,$2.057$\,\,\,\,&$1.976$\\
\hline\hline
\end{tabular}
\end{center}
{\footnotesize Table II: The $v_N\mbox{(simulation)}$, $z_r$, $z_i$,
$A$, $v_N\mbox{(theoretical)}$ and $v_N\mbox{[Eq. (\ref{e3.14})]}$
values for $\tilde\gamma=1$, and $N=10^2$, $10^3$ and $10^5$. The
$v_N\mbox{(theoretical)}$ values are calculated from the theoretical
curves of Fig. \ref{fig7a}.}  \eleq
\vspace{-3mm}
\vspace{-6mm} Notice that as $N$ decreases, according to Table II, the
value of $z_r$ decreases more and more from its $N\rightarrow\infty$
limit $z_0$, while $z_i$ increases. This is an illustration of how the
non-mean field effects become increasingly important behind the tip
region. We should also note two more points about Fig. \ref{fig7a}:
{\it (i)\/} in the absence of any estimate of the correlated diffusion
events for $\Delta t\gtrsim2/v_N$, the theoretical curves lie below
the simulation data (although for $N=10^2$ and $10^3$, it is not so
clearly discernible), and {\it(ii)\/} the agreement between the
theoretical $P(\Delta t)$ curve and the simulation one for $N=10^5$
may appear to be worse than the corresponding ones for $N=10^2$,
$10^3$ and $10^4$, but this may due to the fact that we have had to
continuously cut off particles from the saturation region of the front
on the left to obtain the stochastic simulation data for $N=10^5$
within a reasonable computer time. We have found that the shape of the
$P(\Delta t)$ histogram obtained from the simulation gets slightly
modified depending on how this subtraction in carried out, specially
in the $\Delta t\gtrsim2/v_N$ region.

\begin{figure}[h]
\begin{center}
\includegraphics[width=6.5in,angle=0]{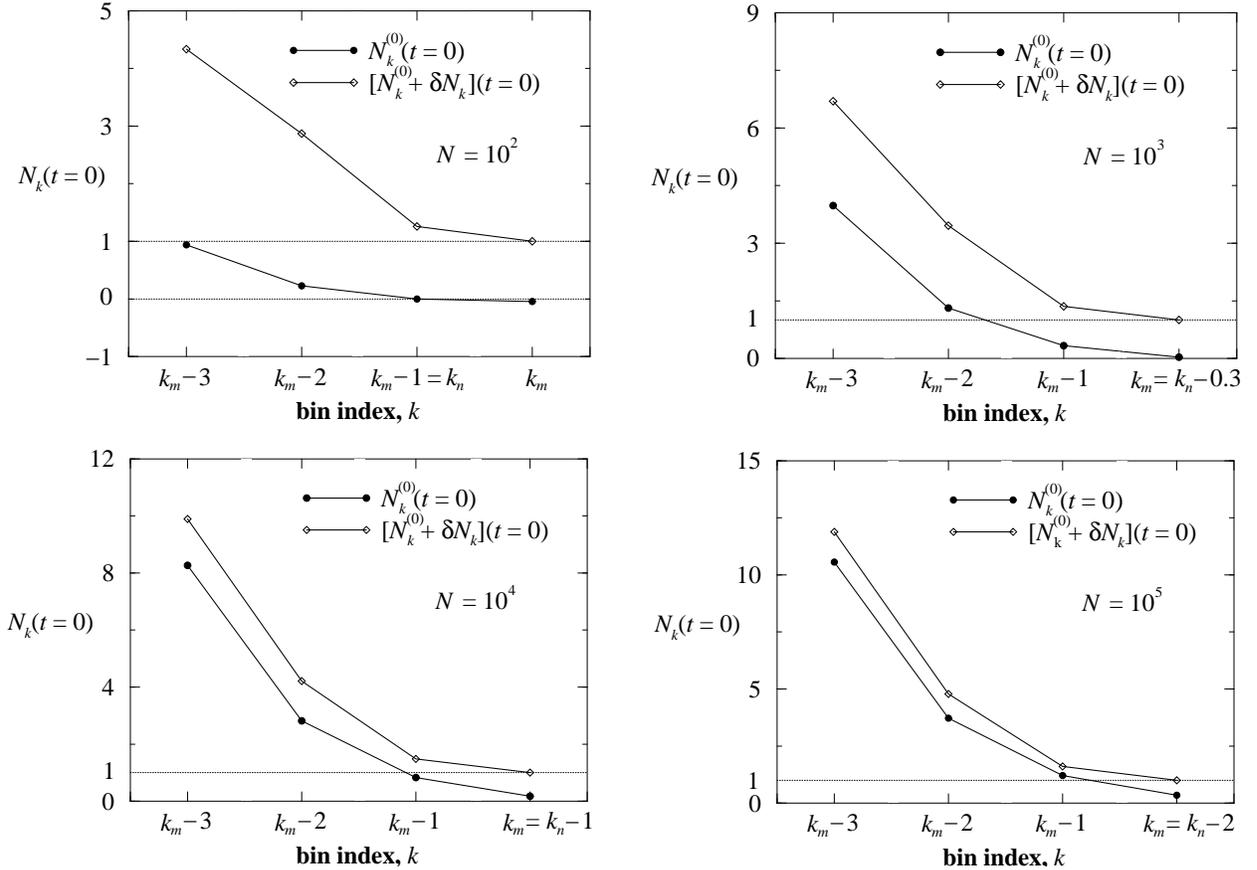}
\end{center}
\caption{Comparison between the $\langle\delta N_k\rangle(t=0)$ and
the $\langle N^{(0)}_k\rangle(t=0)$ values for four foremost bins and
for $N=10^2$, $10^3$, $10^4$ and $10^5$. The angular brackets for the
notations have been omitted in the figure for clarity. Note that as
$N$ increases, the corrections $\langle\delta N_k\rangle(t=0)$ compared
to the $\langle N^{(0)}_k\rangle(t=0)$ profile become less and less
important.\label{comb2345}}   
\end{figure}

We now return to (the issue raised in point {\em 5)\/} of
Sec. \ref{4D}) the increased importance of the stalling effects and
the deviations from the $N\rightarrow\infty$ asymptotic theory for
decreasing values of $N$. Figure \ref{comb2345} shows the comparison
between the {\it actual particle numbers\/}, $\langle\delta
N_k\rangle(t=0)=N\delta\phi_k(t=0)$ and the $\langle
N^{(0)}_k\rangle(t=0)=N\phi^{(0)}_k(t=0)$ values for four foremost
bins, i.e., for $k=k_m$, $k_{m-1}$, $k_{m-2}$ and $k_{m-3}$ (note that
for the sake of clarity, we have omitted the angular brackets for
notations in Fig. \ref{comb2345}). These values have been obtained
self-consistently, while generating the theoretical curves of
Figs. \ref{fig7} and \ref{fig7a}. As expected, it is clear that the
$\delta N_k(t=0)$ values are playing less and less role for increasing
value of $N$. There are couple of more points that one must take
notice of. First, as can be seen from Table II, $A$ increases with
$N$, but only by a small amount. Secondly, it is also clear from
Fig. \ref{comb2345} that w.r.t. the $k_n$-th bin [where $\langle
N^{(0)}\rangle$ vanishes], the position of the $k_m$-th bin (where
$\langle N^{(0)}+\delta N\rangle(t=0)=1$) shifts gradually towards the
left for increasing $N$ (see Table I for the definition of $k_n$). All
these together elucidate that for not very large values of $N$, the
actual $N$ dependence of the front speed $v_N$ is a much more
complicated story than simply the $1/\ln^2N$ relaxation to $v^*$ of
$v_N$. From the trend of the gradual left-shifting of $k_m$
(w.r.t. $k_n$) and the gradual unimportance of the role of $\delta
N_k(t=0)$ values compared to their $N_k(t=0)$, it is conceivable that
for extremely large values of $N$, $k_m\rightarrow k_{m_0}$ and
$\delta N_k(t=0)\rightarrow0$, and it is this limit where the cutoff
(at $\phi^{(0)}=1/N$) picture in Ref. \cite{bd} becomes applicable. In
this sense, the theory of Sec. IV is complementary to that of
Ref. \cite{bd}, as together they span the whole range of $N$ values,
from reasonably small to very large.  \bleq
\vspace{-2mm}
\eleq

\subsection{The Case of $\tilde\gamma=0.1$ and $N=10^4$}

We now investigate the claim made in Sec. IV that the correlated
diffusion events affect the ${\cal P}(\Delta t)$ curve so severely for
low diffusion coefficients that our approach fails badly, by comparing
the theoretical $P(\Delta t)$ curve with the simulation one,
for $\tilde\gamma=0.1$ and $N=10^4$.

\begin{figure}[h]
\begin{center}
\includegraphics[width=2.85in,angle=270]{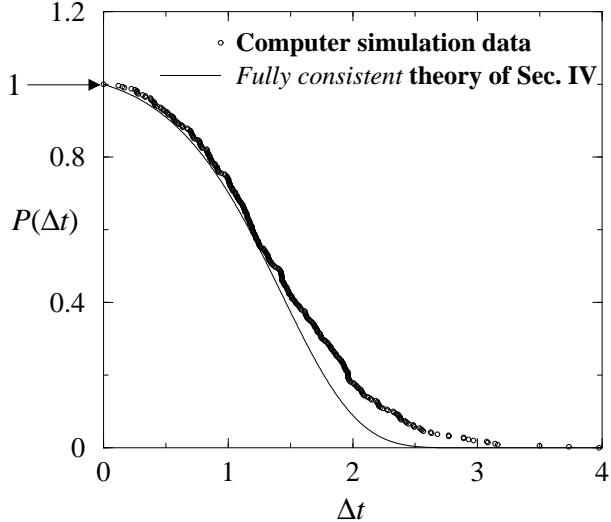}
\caption{The cumulative probability distribution $P(\Delta t)$ as a
function of $\Delta t$ for $N=10^4$ and
$\tilde\gamma=0.1$. \label{fig8}}
\end{center}
\end{figure}

We present the two curves in Fig. \ref{fig8}. The asymptotic speed for
the corresponding pulled front, $v^*$, for this set of parameter
values is given by $v^*=0.7754$ and the simulation results yield
$v_N=0.698$.

The theoretical curve of Fig. \ref{fig8} is analogous to that of
Fig. \ref{fig6} represented by the solid line, and it is obtained by
means of a fully consistent theory of Sec. IV. Notice that the
agreement between the theory and the computer simulation results is
not good beyond $\Delta t\approx1/v_N$. It is also obvious that an
attempt to incorporate the effect of the vacant foremost bin events
does not do any improvements in this case, since the value of
$P(\Delta t)$ is almost zero for $\Delta t\gtrsim2/v_N$. This is very
much expected and a careful examination of the simulation data also
reveals that the vacant foremost events do not occur at all during the
front speed measurement times between $t=200$ and $t=700$ (see
Eq. (\ref{e5.6}) and the paragraph above it). Altogether, this fits 
very nicely in the consistent picture that we have put forward so far,
which simply indicates that the entire discrepancy between the theory
and the computer simulation in Fig. \ref{fig8} is solely due to the
correlated diffusion events.
\vspace{-2mm}

\section{The Case of a Finite Number of Y particles on
the Lattice Sites}\label{knsmodel}

We now briefly turn our attention to the following reaction-diffusion
process X$+$Y $\rightarrow$ 2X on a lattice: at each lattice position,
there exists a bin. Once again, we label the bins by their serial
indices $k$, $k=1,2,3,\ldots,M$, placed from left to right. In the
$k$-th bin, there are a certain number of X particles, denoted by
$N_{\mbox{\tiny X},\,k}$ and a certain number of Y particles, denoted
by $N_{\mbox{\tiny Y},\,k}$. Both $N_{\mbox{\tiny X},\,k}$ and
$N_{\mbox{\tiny Y},\,k}$ are finite. The dynamics of the system is
described by three basic processes: {\em (i)} Diffusion of the X
particles in the $k$-th bin to the ($k-1$)-th or the ($k+1$)-th bin
with a rate of diffusion $\tilde\gamma$. If an X particle in bin 1
jumps towards the left, or an X particle in the $M$-th bin jumps to
the right, then they are immediately replaced. {\em (ii) } Likewise,
diffusion of the Y particles in the $k$-th bin to the ($k-1$)-th or
the ($k+1$)-th bin with a rate of diffusion $\tilde\gamma$. If an X
particle in bin 1 jumps towards the left, or an X particle in the
$M$-th bin jumps to the right, then they are immediately replaced.
{\em (iii)} Reaction to produce an {\it extra} X particle having
annihilated a Y particle (X $+$ Y $\rightarrow$ 2X), with a rate
$1/N$. To study the phenomenon of front propagation for this model,
the initial configuration of the system is taken as $N_{\mbox{\tiny
X},\,k}=N[1-\Theta(k-k_0)]$ and $N_{\mbox{\tiny
Y},\,k}=N\Theta(k-k_0)$ (a step function in the density of the X
particles). Because of the reaction process {\em (iii)}, the number of
X particles in any bin keeps increasing, until the supply of Y
particles in that bin runs out. As a result, the region that is full
of X particles slowly invades the region that is full of Y particles,
and this constitutes a propagating front.

The corresponding equation of the front that is analogous to
Eq. (\ref{e2.5}), is slightly more complicated, and it is given by
\bleq \be \frac{\partial}{\partial t}\langle N_{\mbox{\tiny X},\,k}
(t)\rangle\,=\,\tilde\gamma\,\left[\,\langle N_{{\mbox{\tiny X},\,k}
\,+\,1}(t)\rangle\,+\,\langle N_{{\mbox{\tiny X},\,k}
\,-\,1}(t)\rangle\,-\,2\,\langle N_{\mbox{\tiny X},\,k}
(t)\rangle\,\right]\,+\,\frac{1}{N}\,\left[\langle
N_k(t)\,N_{\mbox{\tiny X},\,k} (t)\rangle\,-\,\langle N^2_{\mbox{\tiny
X},\,k} (t)\rangle\,\right]\,,
\label{e6.1}
\ee \eleq
\noindent where $N_k(t)$ is the total number of particles in the
$k$-th bin at time $t$. In Eq. (\ref{e6.1}), if we replace 
$\langle N_k(t)\,N_{\mbox{\tiny X},\,k} (t)\rangle$ by $\langle
N_k(t)\rangle\,\langle N_{\mbox{\tiny X},\,k} (t)\rangle=N\langle
N_{\mbox{\tiny X},\,k} (t)\rangle$, then one retrieves
Eq. (\ref{e2.5}). In this section, therefore, our purpose is to
investigate if the correlation term $\langle N_k(t)\,N_{\mbox{\tiny
X},\,k} (t)\rangle$ has any bearing on the corrections of the
asymptotic front speed over its corresponding value obtained from the
model analyzed so far.

Front propagation in this model has been studied numerically by
Kessler and co-authors \cite{kns}. Our interest in this model is
motivated by the following observation: in terms of the average number
of X particles in a bin, an appropriate  reaction rate yields an
equation, which is similar to Eq. (\ref{e2.5}). However, the linear
growth term of Eq. (\ref{e2.5}) is replaced by a more complicated
correlation function between the number of X and Y particles in the
$k$-th bin. Nevertheless, near the foremost bin of the X particles,
the number of Y particles is so large that the fluctuations in their
number remains small. Upon neglecting these fluctuations, the linear
growth term for the X particles becomes the same as the one before,
and one therefore expects to the speed correction to stay
unaffected. Our purpose is to check this expectation {\it
numerically\/}, by comparing data for the front speed in this  model
with those given in Eq. (\ref{e5.8}) for Eq. (\ref{e2.5}). The
algorithm that we use in our simulation is similar to that of
Sec. \ref{sec5a}. The value of $M$ is taken to be $1500$ and for the
starting configuration of the system, we use $k_0=50$.

The asymptotic front speed is calculated using
Eqs. (\ref{e5.6}-\ref{e5.7}). The measurement of the front speed
starts at $t=200$ and stops at $t=700$. There are $985$ $\Delta
t$-values in this time interval. The corresponding $v_N(t_j)$
vs. $t_j$ graph is shown below in Fig. \ref{fig10}. Using the same
method of calculation as in Sec. \ref{sec5a}, the asymptotic front
speed for $N=10^4$ and $\tilde\gamma=1$ comes out of the computer 
simulation as \be
v_N\,=\,1.974\pm0.009\,.
\label{e6.5}
\ee

We notice that the error bars of Eqs. (\ref{e5.8}) and (\ref{e6.5})
overlap with each other, and we conclude that the correlations between
the total number of particles and the number of X particles in the
bins [the $\langle N_k(t)\,N_{\mbox{\tiny X},\,k} (t)\rangle$ term in
Eq. (\ref{e6.1})] indeed do not affect the asymptotic front speed, as
we had expected.
\begin{figure}
\begin{center}
\includegraphics[width=2.4in,angle=270]{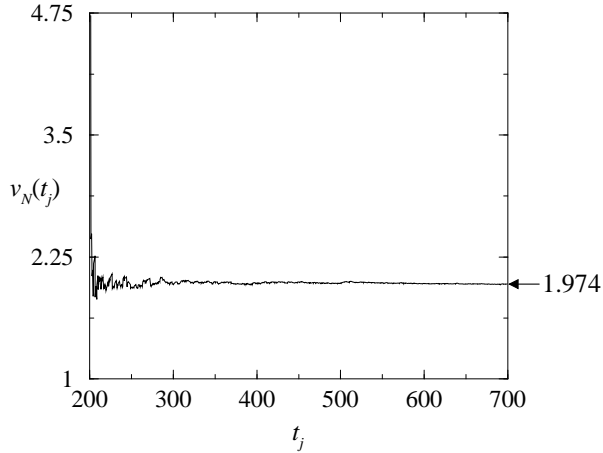}
\caption{Values of $v_N(t_j)$ for $200<t_j<700$ and
$j=1,2,\ldots,980$. As expected, the fluctuations in $v_N(t_j)$ die
out for large values of $j$. \label{fig10}}
\end{center}
\end{figure}

As noted before, Kessler {\em et al.} \cite{kns} claimed that the
prefactor of the speed correction was about a factor 2 different from
the value expected from the asymptotic formula (\ref{e3.14}) of Brunet
and Derrida. Whether this is due to the values of $N$ not being large
enough, or due to their particular way of implementing the stochastic
simulations, is unclear to us. From our data, there is no reason to
believe that for large $N$ there is an essential difference between
the model with finite number of Y particles and the earlier model with
an infinite supply of Y particles, otherwise the asymptotic formula
would be incorrect as $N \to \infty$.

\section{Conclusion and Outlook}
In this paper we have identified a large number of effects that play a
role in the tip region of fluctuating fronts which in the mean-field
limit reduce to pulled fronts. While a full theory from first
principles, which yields explicit predictions for the front speed for
finite $N$ will be hard to come by, we believe that any such theory
will incorporate most of the effects we have analyzed and studied with
computer simulations. One important conclusion from our studies is
that while the asymptotic large-$N$  correction derived by Brunet and
Derrida is universal (in the sense of being independent of the details
of the model) the corrections to this expression  do depend on many
details of the model. In most cases, deviations from the asymptotic
results are significant for values of $N$ that can realistically be
studied.

The message of this paper is as follows: the bulk of a fluctuating
front can still be considered a uniformly translating one and one can
properly define a co-moving co-ordinate system, in which the shape of
the bulk remains unchanged; on the other hand, the position of the tip
of the front in such a co-moving co-ordinate system fluctuates, and
only on average the tip becomes stationary in this co-moving
co-ordinate system. From the mean-field limit of this fluctuating
front, we know that the tip region is very important for its dynamics;
as a result, the fluctuating tip plays a very significant role in
deciding the asymptotic front speed, in which two very important
aspects come to play a role -- discrete nature of particles and
discrete nature of the bin indices. In this paper, we have tried to
formulate a theory to model this fluctuating tip. This theory is still
a mean-field type theory. More specifically, at $t=0$, the shape of
the tip is always the same mean shape, and hence this theory is unable
to capture the correlated diffusion events or the vacant foremost bin
events (although we can estimate the effect of the latter). Any
alternative theory, that one might think at this point, must be able
to take into account these fluctuation effects, which, as previously
explained, must be able to study multi-time correlation functions for
correlated jumps at the tip.

The prospect of such a theory however, looks grim at this point. Not
only the problem becomes highly nonlinear, but also one must realize
that the fluctuations in the number of X particles in the bins near
the tip is of the same order as the number of X particles in them
($\sim1$), and there does not exist any small parameter that one can
do perturbation theory with.

Finally, we note here that we have confined our analysis to cases in
which the growth and hopping terms for few particles per site or bin
are the same as those for a finite density of particles. In such cases
the front speed converges to $v^*$ as $N\to \infty$. One should keep
in  mind, though, that there are also cases where with minor
modifications of the stochastic rules for few particles, one can
arrive at a  situation in which the speed does not  converge to the
pulled speed $v^*$ as $N \to \infty$, even though in the mean field
limit one obtains a dynamical equation that admits pulled fronts. We
will discuss this in more detail elsewhere \cite{deb2}.

ACKNOWLEDGEMENT: D. P. wishes to acknowledge many important,
motivating and stimulating discussions with Hans van Leeuwen, Ramses
van Zon and Zohar Nussinov, the help from Ellak Somfai on setting up
the computer simulations used to obtain results in this paper, and the
financial support from ``Fundamenteel Onderzoek der Materie'' (FOM).

\appendix
\section{Summary of the derivation of the generalized velocity correction formula}\label{bdextension}
In this appendix, we derive the generalized velocity correction
formula, Eq.~(\ref{e3.12}) and its interpretation.

Without any loss of generality, we can express the front solution
$\phi(\xi)$ for $v_N < v^*$ by (Cf. \cite{bd,kns,levine,vs2,ebert})
\be \phi(\xi)\,=\,A\,\sin\,[\,z_i\,\xi\,+\,\beta\,]\,\exp(-\,z_r\,\xi)
\label{e3.3app}
\ee      at the leading edge of the front, where $z_r=\mbox{Re}(z)$
and $z_i=\mbox{Im}(z)$. The corresponding dispersion relation is then
given by        \be
z_r\,v_N&=&2\,\tilde\gamma\,(\cosh\,z_r\,\cos\,z_i\,-\,1)\,+\,1\quad\mbox{and}\nonumber\\&&z_i\,v_N\,=\,2\,\tilde\gamma\,\sinh\,z_r\,\sin\,z_i\,.
\label{e3.4}
\ee       The additive phase $\beta$ in Eq. (\ref{e3.3}) can be scaled
away by redefining $A$ and the position of the origin from where $\xi$
is measured. We therefore drop $\beta$  in this appendix. Since the
scaled particle density has to be positive, i.e., $\phi(\xi)\geq0$,
the physical linear solution regime must be confined within the range
where $0<z_i\,\xi\leq\pi$. We now make a notational choice to denote
the comoving coordinate corresponding to the node of the sine function
in Eq. (\ref{e3.3}), by $\xi_c+1$, i.e.,     \be
z_i\,(\xi_c\,+\,1)\,=\,\pi\,.
\label{e3.5}
\ee      {\it One should understand at this point that although
Eq. (\ref{e3.3app}) suggests that there is a second node of
$\phi(\xi)$, where the argument of the $\sin$ function becomes zero,
such a node does not exists. Much before the argument of the $\sin$
function becomes zero, the nonlinear saturation term becomes important
and the solution (\ref{e3.3app}) for the linearized equation does not
hold any longer}. In this overly simplified mean-field description,
$\xi_c$ plays the role of the comoving coordinate of the foremost
bin. The mean field description of the front is then completed by
claiming that $\phi(\xi)=0$ for $\xi\geq\xi_c+1$. Let us also denote
the density of the X particles in the ``foremost bin'', which in this
approximation is at $\xi_c$, by $a/N$, to have    \be
A\,\sin\,[\,z_i\,\xi_c\,]\,\exp(-\,z_r\,\xi_c)\,=\,\frac{a}{N}\,.
\label{e3.6}
\ee

Once the parameters $A$ and $a$ are known,
Eqs. (\ref{e3.4}-\ref{e3.6})  form a set of four equations  for four
unknowns, $z_r$, $z_i$, $\xi_c$  and $v_N$, which we can then solve
numerically for the asymptotic front  speed $v_N$.

In order to put our results in a particular form that facilitates
comparison with the earlier results in literature \cite{bd}, we
analyze  Eqs. (\ref{e3.4}-\ref{e3.6}) for large $N$.  First, with the
help of the Eq. (\ref{e3.5}), we reduce Eq. (\ref{e3.6}) to    \be
A\,\sin\,z_i\,\exp(-\,z_r\,\xi_c)\,=\,\frac{a}{N}\,.
\label{e3.7}
\ee      Next, having introduced a new variable $\mu$, such that \be
\xi_c\,=\,\frac{\ln\,N}{z_r}\,+\,\mu\,,
\label{e3.8}
\ee      and using Eq. (\ref{e3.5}), Eq. (\ref{e3.7}) is further
reduced to an implicit equation in $\mu$:    \be
z_r\,\mu\,=\,\left[\,\ln\,\frac{A}{a}\,+\,\ln\,\left\{\sin\frac{\pi\,z_r}{\ln\,N\,+\,1\,+\,\mu}\,\right\}\right]\,.
\label{e3.9}
\ee      Since $N$ is large, one can solve this implicit equation
$\mu$ by means of a simple successive approximation procedure. At the
lowest order, one can drop the $\mu$ term in the denominator of the
argument of the sine function in Eq. (\ref{e3.9}) and obtain      \be
\mu\,\approx\,\frac{1}{z_r}\,\left[\,\ln\,\frac{A}{a}\,+\,\ln\,\left\{\sin\frac{\pi\,z_r}{\ln\,N\,+\,1}\,\right\}\right]\,.
\label{e3.10}
\ee      Finally, $z_i$ can be obtained from Eqs. (\ref{e3.5}),
(\ref{e3.8})  and (\ref{e3.10}) as      \be
z_i\,\approx\,\frac{\pi\,z_r}{\ln\,N\,+\,z_r\,+\,\displaystyle{\ln\,\frac{A}{a}\,+\,\ln\,\left[\sin\frac{\pi\,z_r}{\ln\,N\,+\,1}\,\right]}}\,.
\label{e3.11}
\ee

By now, we have eliminated the unknown $\xi_c$ and reduced the problem
to solving three unknowns, $z_r$, $z_i$ and $v_N$ from three
equations, Eqs. (\ref{e3.4}) and (\ref{e3.11}). From
Eq. (\ref{e3.11}), one can see that for large $N$, the approach of
$z_i$ to zero is extremely slow, going only as $\ln^{-1}N$ and also
the fact that for the strict limit of infinite $N$, $z_i=0$, which
reduces Eq. (\ref{e3.4}) to Eq. (\ref{e3.1}), as it should. For large
$N$, therefore, one expects that $z_r\approx\,z_0$ and $z_i\ll1$, and
one can expand $v_N$ around its value for $z=z_0$.  Upon expanding
$v_N$ around $v^*$, and using the solution of $z_i$ from
Eq. (\ref{e3.11}) with $z_r$ replaced by $z_0$, we then find that the
asymptotic speed $v_N$ is given by Eq.~(\ref{e3.12}).

\section{Determination of $A$}\label{Aappendix}

Solving the full nonlinear difference-differential equation,
Eq. (\ref{e2.6}), is not an easy task. For a given set of values of
$\tilde\gamma$ and $N$, there are essentially two methods to determine
the value of $A$. The first one is to obtain the solution close to the
saturation value $\phi=1$ and thereafter iterate the solution until
one reaches $\phi\approx0$. Close to the saturation value $\phi=1$,
one can define $\eta(\xi)=1-\phi(\xi)$, which reduces Eq. (\ref{e2.6})
to an equation in $\eta(\xi)$, given by  \be
-\,v_N\,\frac{d\eta}{d\xi}\,=\,\tilde\gamma\,[\,\eta(\xi+1)\,+\,\eta(\xi-1)\,-\,2\eta(\xi)\,]\nonumber\\&&\hspace{-3cm}-\,\eta(\xi)\,+\,\eta^2(\xi)\,.
\label{e5.9}
\ee  For $\eta$ values close to zero, the solution of Eq. (\ref{e5.7})
is given by the linearized equation   \be
-v_N\frac{d\eta}{d\xi}=\tilde\gamma[\eta(\xi+1)+\eta(\xi-1)-2\eta(\xi)]-\eta(\xi)\,,
\label{e5.10}
\ee  with the corresponding solution
$\eta(\xi)=B_1\exp[z_l(\xi-\xi_0)]$.  Substitution of this solution in
Eq. (\ref{e5.10}) yields the dispersion relation between $v_N$ and
$z_l$  \be  -\,v_Nz_l\,=\,2\gamma\,[\,\cosh\,z_l\,-\,1\,]\,-\,1\,.
\label{e5.11}
\ee  One can then iterate this solution towards $\phi=0$. The full
solution of Eq. (\ref{e5.9}) can be written as  \be
\eta(\xi)\,=\,\sum_{n=1}^{\infty}\,B_n\,\exp[\,nz_l\,(\xi-\xi_0)]\,,
\label{e5.12}
\ee  where the corresponding $B_n$ values are obtained from the
recursion relation  \be
B_n\,=\,\frac{B_1B_{n-1}\,+\,B_2B_{n-2}\,+\,\ldots\,+\,B_{n-1}B_1}{1\,-\,nv_Nz_l\,-\,2\tilde\gamma\,(\cosh\,nz_l\,-\,1)}\,.
\label{e5.13}
\ee  As a starting point for constructing the solution near $\phi=1$,
one can choose arbitrary values of $\xi_0$ and $B_1$, so long as the
value of $B_1$ is sufficiently small. At small values of $B_1$, any
scaling of $B_1$ amounts to a simple shift  of the origin
$\xi_0$. Finally, one can then match the solution, thus obtained, to
the form $\phi(\xi)=A\sin[z_i(\xi-\xi_1)]\exp[z_r(\xi-\xi_1)]$ near
$\phi=0$ and determine the value of $A$.

The second method to obtain the numerical value of $A$ is to assume a
certain value of $A$ close to $\phi=0$ with the functional form
$\phi(\xi)=A\sin[z_i(\xi-\xi_1)]\exp[z_r(\xi-\xi_1)]$ and then
continue to iterate the corresponding solution in the direction of
$\phi=1$ in a similar manner. This time, if the assumed value of $A$
is correct, then close to $\phi=1$, one must recover the exponential
behaviour of $\phi(\xi)$, as in Eq. (\ref{e5.12}). However,  we have
found that the first method is stable under small changes in the
starting value of $B_1$, while the second method is not stable under
small changes in the assumed value of $A$.

The first method should therefore be the natural choice, albeit from a
practical point of view, one needs a very large number of $B_n$ values
to extend the solution of $\phi(\xi)$ all the way up to $\phi=0$. In
practice, we have therefore used a ``double shooting'' method
\cite{press}, in which the functions are calculated from both sides,
and then matched somewhere in the middle.

The matching of the values of the functions and their derivatives at
$\phi_0$ requires the values of $z_l$, $z_r$ and $z_i$ to be
determined externally. For $\tilde\gamma=1$ and $N=10^4$, the values
of $z_l$, $z_r$, $z_i$ and $A$ are numerically obtained from
Eqs. (\ref{e5.8}), (\ref{e3.4}) and (\ref{e5.11}) as  \be
z_l\,=\,0.4187\,\mp\,0.0008\,,\nonumber\\&&\hspace{-3.56cm}z_r\,=\,0.877\,\pm\,0.002\,,\nonumber\\&&\hspace{-3.56cm}z_i\,=\,0.264\,\mp\,0.007\quad\mbox{and}
\nonumber\\&&
\hspace{-3.56cm}A\,=\,0.961\,\pm\,0.012\,.
\label{e5.14}
\ee  Of course, the numerical value of $A$ depends on the origin,
where from $\xi$ is measured for the form in Eq. (\ref{e3.3}). In
Eq. (\ref{e5.14}) above, the value of $A$ is determined with
$\beta=0$, i.e., the solution of the linearized equation at the
leading edge of the front is zero at $\xi=\pi/z_i$ . We mention here
that the uncertainty in these numbers is determined by the uncertainty
in $v_N$, not by the inaccuracy of the numerical method.
 
%\end{appendix}

\ecols


\begin{thebibliography}{99}
\bibitem{notepulled} One should be careful in considering the limits
in discussing this issue. Strictly speaking, for any finite particle
number $N$ a front is not pulled, but weakly pushed. Hence for any
finite $N$ the true asymptotic scaling should be of the KPZ type.
\bibitem{notepushed} For the F-KPP model with a cutoff in the growth
term that Brunet and Derrida \cite{bd} introduced to model the finite
particle effects, it is easy to convince oneself explicitly that the
fronts are in fact always weakly pushed, according to the
classification of \cite{ebert}.
\bibitem{bdrealized} Having realized this, Brunet and Derrida modified
the corresponding mean field equation analogous to Eq. (\ref{e3.0}) to
take it into account \cite{bd}. We make detailed comments about it at
the end of this section.
\bibitem{notematching} We note that several aspects of the analysis in
this section show similarities with elements of the analysis of
\cite{bd}. However, their analysis is simpler because they introduced
a cutoff in a phenomenological way into the growth rate in the
continuum mean field equation. As a result, they could simply  match
the two regions where the growth rate was different. Our analysis is
beset by the fact that we do attempt to take the lattice effects into
account (this already makes matching impossible), as well as the
stalling effects and the changes this gives in the profile behind the
foremost bin.
\bibitem{timescale} In actuality, at the instant step (b) is over, the
population of the  $(k_m-1)$-th bin has grown to value bigger than
1. As a result, the diffusion of an X particle from the $(k_m-1)$-th
bin to the $k_m$-th bin takes slightly smaller time than the time
scale for a new foremost bin creation $1/v_N$. A $1/v_N$ time scale
would be applicable if the population of the  $(k_m-1)$-th bin is
exactly 1 at the instant step (b) is over. We choose to ignore this,
and by choosing to do so, we overestimate the value of $t_0$ by a
small amount.
\bibitem{2/v_N} Of course, in an actual process, $t_0$ is determined
probabilistically and not in such a sharp manner. However, we should
remember that this is only an estimate, and we will show later that it
works reasonably well.
\bibitem{highd} For large values of $\tilde\gamma$, there can be
another kind of event, where the vacant foremost bin event takes place
more than once for the same foremost bin. This is more likely for
large values of $\tilde\gamma$, since the hops from the foremost bin
towards the left is much more likely before the number of X particles
gets a chance to grow in the foremost bin. In that case, one can
simply extend this existing way of estimating $t_0$ to
$t_0\approx4/v_N$, $t_0\approx6/v_N$, $\ldots$ etc.
\bibitem{remember} We must not use this value of $a$ to calculate the
speed correction $v^*-v_N$, since it is not an ``externally determined
quantity''. If we do, then it would be a case of obtaining the
solution of an equation, where the solution itself is used to solve
the equation, and hence a tautology.
\bibitem{linear} We have verified that at $k_b$, the front shape is
still given by the solution of the linear equation, Eq. (\ref{e4.6}).
\bibitem{recursive} This means that we start with a guess value of
$k_m$, obtain the theoretical ${\cal P}(\Delta t)$ curve, compare it
with the ${\cal P}(\Delta t)$ curve obtained from the computer
simulations, and use the nature of discrepancy to make the next guess
value of $k_m$. This process is repeated until one converges to the
value of $k_m$, for which the best agreement between the theoretical
and the simulation ${\cal P}(\Delta t)$ curves is achieved. In this
recursive feedback method, specifically for $\tilde\gamma\approx1$ and
$\tilde\gamma\ll1$, one should remember to compare the theoretical and
the simulation ${\cal P}(\Delta t)$ curves for $0<\Delta
t\lesssim1/v_N$ while obtaining the numerical value of $k_m$, since
for $\Delta t\sim2/v_N$, there are other complications that come to
play a very significant role. These effects are discussed in
Sec. \ref{sec4c}.
\end{thebibliography}
\end{document}